\documentclass[a4paper,11pt]{article}
\usepackage{epsfig}
\usepackage{graphicx}
\usepackage{dcolumn}
\usepackage{bm}
\usepackage{longtable}
\usepackage{amssymb}
\usepackage{amsfonts}
\usepackage{amsmath}
\usepackage{hyperref}
\usepackage{slashed}

\allowdisplaybreaks[1]

\newcommand{\be}{\begin{equation}}
\newcommand{\ee}{\end{equation}}
\newcommand{\ba}{\begin{align}}
\newcommand{\ea}{\end{align}}
\newcommand{\one}{{\rm 1\kern -.9mm l}}

\newcommand{\U}{U}

\def\m{\mu}

\makeatletter

\@addtoreset{equation}{section}
\makeatother
\begin{document}

\begin{titlepage}

     \thispagestyle{empty}
    \begin{flushright}
        \hfill {DFPD-14/TH/04} \\

    \end{flushright}


    \begin{center}

         {\huge{\textbf{No Fermionic Wigs \\\vspace{3pt} for BPS Attractors in 5 Dimensions}\\}}
         \vspace{20pt}

{\Large{{\bf Lorenzo G. C.~Gentile$^{1,2}$}, {\bf Pietro A. Grassi$^{1,3}$}, \\{\bf Alessio Marrani$^{4}$}, {\bf Andrea Mezzalira$^{5}$}, and {\bf Wafic A. Sabra$^{6}$}}}

        \vspace{20pt}

         {$^1$ DISIT, Universit\`{a} del Piemonte Orientale, \\via T. Michel, 11, Alessandria, I-15120, Italy,\\
        \texttt{pgrassi@mfn.unipmn.it}}

         \vspace{5pt}

         {$^2$ Dipartimento di Fisica �\U{201c}Galileo Galilei�\U{201d},\\Universit\`a di Padova, via Marzolo 8, I-35131 Padova, Italy\\
         and INFN, Sezione di Padova, via Marzolo 8, I-35131, Padova, Italy,\\
        \texttt{lgentile@pd.infn.it}}

         \vspace{5pt}

        {$^3$ INFN - Gruppo Collegato di Alessandria - Sezione di Torino}

         \vspace{5pt}

{$^4$ Instituut voor Theoretische Fysica, KU Leuven,\\
Celestijnenlaan 200D, B-3001 Leuven, Belgium,\\
\texttt{alessio.marrani@fys.kuleuven.be}}

\vspace{5pt}

{$^5$ Physique Th\'eorique et Math\'ematique,\\
Universit\'e Libre de Bruxelles, C.P. 231, B-1050 Bruxelles, Belgium,
\\\texttt{andrea.mezzalira@ulb.ac.be} }

\vspace{5pt}

{$^6$ Centre for Advanced Mathematical Sciences and Physics Department,\\
American University of Beirut, Lebanon,
\\\texttt{ws00@aub.edu.lb} }

 \vspace{5pt}

\end{center}

\vfill{}

\begin{abstract}
{\vspace{.3cm}

       \noindent
We analyze the {\it fermionic wigging} of $1/2$--BPS (electric) extremal black hole attractors in $\mathcal{N}=2$,
$D=5$ ungauged Maxwell--Einstein supergravity theories, by exploiting anti--Killing spinors supersymmetry transformations.
Regardless of the specific data of the real special geometry of the manifold defining the scalars of the vector multiplets, and differently from the $D=4$ case,
we find that there are {\it no corrections} for the near--horizon attractor value of the scalar fields;
an analogous result also holds for $1/2$--BPS (magnetic) extremal black string.
Thus, the attractor mechanism receives {\it no fermionic corrections} in $D=5$ (at least in the BPS sector).

}
\end{abstract}
\vfill{}
\vspace{1.5cm}
\end{titlepage}

\vfill
\eject

\setcounter{footnote}{0}

\tableofcontents


\section{\label{Intro}Introduction}

The question concerning the presence or absence of hairs of any kind around
a black hole is very compelling and, of course, it has been studied from
several points of view. Nonetheless, recently some of the authors of the
present work re-posed the question by considering possible fermionic hairs
(first in \cite{Burrington:2004hf}, and then in a series of papers \cite%
{Gentile:2011jt}) for non-extremal, as well as BPS black holes. The first
paper on the subject is due to Aichelburg and Embacher \cite%
{Aichelburg:1986wv}. They considered asymptotically flat black hole solution
in $\mathcal{N}=2$, $D=4$ supergravity without vector multiplets and
computed iteratively the supersymmetric variations of the background in
terms of the flat-space Killing spinors. In that paper, they were able to
compute some of the physical quantities such as the corrections to the
angular momentum, while other interesting properties cannot be seen at that
order of the expansion. Afterwards, the works \cite{Burrington:2004hf} and
\cite{KL-1} applied their technique to some examples of BPS black hole, up
to the fourth order in the supersymmetry transformation.

In particular, for extremal black hole solutions, the \textit{attractor
mechanism} \cite{AM-Refs} is a very interesting and important physical
property; essentially, it states that the solution at the horizon depends
only on the conserved charges of the system, and is independent of the value
of the matter fields at infinity. This is related to the \textit{no-hair
theorem}, under which, for example, a BPS black hole solution depends only
upon its mass, its angular momentum and other conserved charges. As said,
the authors of \cite{KL-1} addressed the question whether the attractor
mechanism has to be modified in the presence of fermions. The conclusion was
that, at the level of approximation of their computations, in the case of
double-extremal BPS solutions, the mechanism is unchanged. In \cite%
{Burrington:2004hf} $\mathcal{N}=2$, $D=5$ AdS black holes were
investigated, and it was found that the solution, as well as its asymptotic
charges, get modified at the second order due to fermionic contributions.
However, in \cite{Burrington:2004hf} the attractor mechanism and its
possible modifications was not considered.

In \cite{Gentile:2013gea}, the fermionic wig for asymptotically flat BPS
black holes in $\mathcal{N}=2$, $D=4$ supergravity coupled to matter was
investigated. There, it has been shown that the attractor mechanism gets
modified at the fourth order even in the case of double extremal solutions
in the simplest example of $\mathcal{N}=2$ supergravity coupled to a single
matter field (\textit{minimally coupled} vector multiplet). The surprising
result is that to the lower orders all corrections vanish for the BPS
solution, while at the fourth order, despite several cancellations due to
special geometry identities, some terms do survive, and thus the attractor
gets modified.

It has also been noticed that there are situations in which some
combinations of charges render the attractor modifications null; this led to
the conjecture that, in those $D=4$ models admitting an uplift to $5$
dimensions, the attractor mechanism is unmodified by the fermionic wig. That
motivated us to study in full generality the $D=5$ case, by means of the
same techniques; we found that \textit{there is no modification to the
attractor mechanism up to forth order} for all the ungauged $\mathcal{N}=2$,
$D=5$ supergravity models coupled to vector multiplets. This is a rather
strong result, and it has been obtained for a \textit{generic} real special
geometry of the manifold defined by the scalars of the vector multiplets.
The cancellations appear to be due to identities of the special geometry, as
well as to the extremal black hole solutions taken into account (\textit{cfr.%
} Eq. (5.1)).

We should point out that the wigging is computed by performing a
perturbation of the unwigged purely bosonic BPS extremal black hole solution
keeping the radius of the event horizon unchanged. The complete analysis,
including the study of the fully-backreacted wigged black hole metric, will
be presented elsewhere.

\bigskip

The plan of the paper is as follows. In Sec. \ref{D=5-Ungauged} we recall
some basics of $\mathcal{N}=2$, $D=5$ ungauged Maxwell-Einstein
supergravity. The fermionic wigging is then presented in Sec. \ref%
{Fermi-Wigging}, and its evaluation on the purely bosonic background of an
extremal BPS black hole is performed in Sec. \ref{Purely-Bosonic}. The
near-horizon conditions are applied in Sec. \ref{Wig-BPS-D=5}, obtaining the
\textit{universal} result of vanishing wig corrections to the attractor
value of the scalar fields of the vector multiplets in the near-horizon
geometry. The \textit{universality} of this result resides in its \textit{%
independence} on the data of the real special geometry endowing the scalar
manifold of the supergravity theory. Comments on this result and further
remarks and future directions are given in Sec. \ref{Conclusion}.

Three Appendices, specifying notations and containing technical details on
the wigging procedure, are presented.

\section{ \label{D=5-Ungauged}Ungauged $\mathcal{N}=2$, $D=5$ MESGT}

Following \cite{Gunaydin:1983bi}--\nocite{Ceresole:2000jd}--\cite%
{Bergshoeff:2004kh}, we consider $\mathcal{N}=2$, $D=5$ \textit{ungauged}
Maxwell--Einstein supergravity theory (MESGT), in which the $\mathcal{N}=2$
gravity multiplet $\left\{ e_{\mu }^{a},~\psi _{\mu }^{i},~A_{\mu }\right\} $
is coupled to $n_{V}$ Abelian vector multiplets\footnote{$i=1,2$ of the
fundamental $\mathbf{2}$ of $USp(2)\sim SU(2)$ $\mathcal{R}$--symmetry, $%
x=1,\dots ,n_{V}$ and $I=0,1,\dots ,n_{V}$, where the 0 index pertains to
the $D=5$ graviphoton. Note that $\gamma _{\mu }$ denote the $D=5$ gamma
matrices. Moreover, we adopt the convention $\kappa =1$ (\textit{cfr. e.g.}
App.~C of \cite{Bergshoeff:2004kh}).} $\left\{ A_{\mu },\lambda ^{xi},\phi
^{x}\right\} $, with neither hyper nor tensor multiplets\footnote{%
When not indicated, spinor indices are contracted using the standard $%
SU\left( 2\right) $ metric $\varepsilon ^{ij}$ (see appendix \ref%
{App-Notations-Ids}).} :
\begin{subequations}
\begin{align}
\delta e_{\mu }{}^{a}=& {\textstyle\frac{1}{2}}\bar{\epsilon}\gamma ^{a}\psi
_{\mu }\,, \\
\delta \psi _{\mu }^{i}=& D_{\mu }(\hat{\omega})\epsilon ^{i}+{\textstyle%
\frac{\mathrm{i}}{4\sqrt{6}}}h_{I}\tilde{F}_{\nu \rho }^{I}\left( {\gamma }%
_{\mu }^{\phantom{\mu}\nu \rho }-4\delta _{\mu }^{\nu }{\gamma }^{\rho
}\right) \epsilon ^{i}+  \notag \\
& -\,\frac{1}{6}\epsilon _{j}\bar{\lambda}^{ix}\gamma _{\mu }\lambda
_{x}^{j}+\frac{1}{12}\gamma _{\mu \nu }\epsilon _{j}\bar{\lambda}^{ix}\gamma
^{\nu }\lambda _{x}^{j}+  \notag \\
& -\frac{1}{48}\gamma _{\mu \nu \rho }\epsilon _{j}\bar{\lambda}^{ix}\gamma
^{\nu \rho }\lambda _{x}^{j}+\frac{1}{12}\gamma ^{\nu }\epsilon _{j}\bar{%
\lambda}^{ix}\gamma _{\mu \nu }\lambda _{x}^{j}\,, \\
\delta h^{I}=& -\frac{1}{\sqrt{6}}\mathrm{i}\bar{\epsilon}\lambda
^{x}h_{x}^{I}\,, \\
\delta \phi ^{x}& ={\textstyle\frac{1}{2}}\mathrm{i}\bar{\epsilon}\lambda
^{x}\,, \\
\delta A_{\mu }^{I}=& -\frac{1}{2}\bar{\epsilon}\gamma _{\mu }\lambda
^{x}h_{x}^{I}-\frac{\sqrt{6}}{4}ih^{I}\bar{\epsilon}\psi _{\mu }\,, \\
\delta \lambda ^{xi}=& -{\textstyle\frac{\mathrm{i}}{2}}\widehat{%
\slashed{\mathcal D}}\phi ^{x}\epsilon ^{i}-\delta \phi ^{y}\Gamma
_{yz}^{x}\lambda ^{zi}+{\textstyle\frac{1}{4}}\gamma \cdot \tilde{F}%
^{I}h_{I}^{x}\epsilon ^{i}+  \notag \\
& +{\textstyle\frac{1}{4\sqrt{6}}}T^{xyz}\left[ 3\epsilon _{j}\bar{\lambda}%
_{y}^{i}\lambda _{z}^{j}-\gamma ^{\mu }\epsilon _{j}\bar{\lambda}_{y}^{i}{%
\gamma }_{\mu }\lambda _{z}^{j}-{\textstyle\frac{1}{2}}\gamma ^{\mu \nu
}\epsilon _{j}\bar{\lambda}_{y}^{i}{\gamma }_{\mu \nu }\lambda _{z}^{j}%
\right] \,,
\end{align}%
where
\end{subequations}
\begin{subequations}
\begin{align}
\mathcal{F}_{\mu \nu }^{I}=& 2\partial _{\left[ \mu \right. }A_{\left. \nu %
\right] }^{I}\,,  \label{jazz1} \\
\tilde{F}_{\mu \nu }^{I}=& \mathcal{F}_{\mu \nu }^{I}+\bar{\psi}_{\left[ \mu
\right. }\gamma _{\left. \nu \right] }\lambda ^{x}h_{x}^{I}+\frac{i\sqrt{6}}{%
4}\bar{\psi}_{\mu }\psi _{\nu }h^{I}\,, \\
T_{xyz}=& C_{IJK}h_{x}^{I}h_{y}^{J}h_{z}^{K}\,, \\
\Gamma _{xy}^{w}=& h_{I}^{w}h_{x,y}^{I}+\sqrt{\frac{2}{3}}T_{xyz}g^{zw}\ .
\end{align}%
From the \text{Vielbein} postulate, the $\mathcal{N}=2$ spin connection
reads
\end{subequations}
\begin{equation}
{\hat{\omega}_{\mu }^{ab}}=\frac{1}{2}{e_{c\mu }}\left[ \Omega ^{abc}-\Omega
^{bca}-\Omega ^{cab}\right] +K_{\phantom{a}\mu }^{a\phantom{\mu}b}\,,
\end{equation}%
where 
${\Omega ^{abc}:}=e^{\mu a}e^{\nu b}\left( \partial _{\mu }{e_{\nu }^{c}}%
-\partial _{\nu }{e_{\mu }^{c}}\right) 
$ 
and 
$K_{\phantom{a}\mu }^{a\phantom{\mu}b}:=-\frac{1}{2}\bar{\psi}^{\left[
b\right. }\gamma ^{\left. a\right] }\psi _{\mu }-\frac{1}{4}\bar{\psi}%
^{b}\gamma _{\mu }\psi ^{a}$. 
The covariant derivatives are defined as
\begin{subequations}
\begin{align}
\widehat{\mathcal{D}}_{\mu }\phi ^{x}& =\partial _{\mu }\phi ^{x}-{\textstyle%
\frac{1}{2}}\,\mathrm{i}\bar{\psi}_{\mu }\lambda ^{x}\,, \\
\mathcal{D}_{\mu }h^{I}=& \partial _{\mu }h^{I}=-\sqrt{{\textstyle\frac{2}{3}%
}}h_{x}^{I}\partial _{\mu }\phi ^{x}=-\sqrt{{\textstyle\frac{2}{3}}}h_{x}^{I}%
\mathcal{D}_{\mu }\phi ^{x}\,, \\
\mathcal{D}_{\mu }\lambda ^{xi}=& \partial _{\mu }\lambda ^{xi}+\partial
_{\mu }\phi ^{y}\Gamma _{yz}^{x}\lambda ^{zi}+{\textstyle\frac{1}{4}}\omega
_{\mu }{}^{ab}\gamma _{ab}\lambda ^{xi}\,, \\
\mathcal{D}_{\mu }\psi _{\nu }^{i}=& \left( \partial _{\mu }+{\textstyle%
\frac{1}{4}}\omega _{\mu }{}^{ab}\gamma _{ab}\right) \psi _{\nu }^{i}\,,
\end{align}%
and (\cite{Gunaydin:1983bi}; see also \textit{e.g.} Eq.~(C.10) of \cite%
{Bergshoeff:2004kh})
\end{subequations}
\begin{subequations}
\label{covDer2}
\begin{align}
\nabla _{y}h_{x}^{I}& =-\sqrt{\frac{2}{3}}\left(
h^{I}g_{xy}+T_{xyz}h^{Iz}\right) \,, \\
\nabla _{y}h_{Ix}& =\sqrt{\frac{2}{3}}\left(
h_{I}g_{xy}+T_{xyz}h_{I}^{z}\right) \ .
\end{align}%
Note that only $\omega _{\mu }^{ab}$ (and not $\hat{\omega}_{\mu }^{ab}$)
occurs in the covariant derivative of the gravitino. Furthermore, it holds
that\footnote{%
In the present treatment, $C_{IJK}$ denotes the $\mathcal{C}_{IJK}$ of \cite%
{VanProeyen:1999ni}, their difference being just a rescaling factor.} (see
also \textit{e.g.} \cite{Ferrara:2006xx,Ceresole:2007rq,Cerchiai:2010xv})
\end{subequations}
\begin{subequations}
\label{hh1}
\begin{align}
h_{x}^{I}\equiv & -\sqrt{\frac{3}{2}}\partial _{x}h^{I},\qquad h_{Ix}\equiv
a_{IJ}h_{x}^{J}\,, \\
a_{IJ}=& -2C_{IJK}h^{K}+3h_{I}h_{J}\,, \\
C_{IJK}h^{I}h^{J}h^{K}=& 1,\qquad h_{I}h^{I}=1\ .
\end{align}%
It is worth pointing out that in $D=5$ Lorentzian signature no chirality is
allowed, and the smallest spinor representation of the Lorentz group is
given by symplectic Majorana spinors; for further details, see App.~\ref%
{App-Notations-Ids}.


\section{\label{Fermi-Wigging}Fermionic Wigging}

We now proceed to perform the \textit{fermionc wigging}, by iterating the
supersymmetry transformations of the various fields generated by the \textit{%
anti-Killing spinor} $\epsilon $ (for a detailed treatment and further
details, \textit{cfr. e.g.} \cite{Gentile:2012jm,Gentile:2013gea});
schematically denoting all wigged fields as $\widehat{\Phi }$ and the
original bosonic configuration by $\Phi$, the following expansion holds:

\end{subequations}
\begin{equation}
\widehat{\Phi }=e^{\delta }\Phi =\Phi +\delta \Phi +{\frac{1}{2}}\delta
^{(2)}\Phi +{\frac{1}{3!}}\delta ^{(3)}\Phi +{\frac{1}{4!}}\delta ^{(4)}\Phi
\,,  \label{expansion0}
\end{equation}%
where, as in \cite{Aichelburg:1986wv}, the expansion truncates at the fourth
order because of the $4$-Grassmannian degrees of freedom that $\epsilon $
contains.\footnote{%
In the present paper we will deal with a BPS background so just half of the
supersymmetries are preserved.}

\subsection{\label{2nd-Order}Second Order}

In order to give an idea on the structure of the iterated supersymmetry
transformations on the massless spectrum of the theory under consideration,
we present below the second order transformation rules\footnote{%
By exploiting Eq.~(3.16) of \cite{Cerchiai:2010xv}, both $\nabla _{t}T^{xyz}$
and $\nabla _{t}\Gamma _{yz}^{x}$ can be related to the covariant derivative
of the Riemann tensor $R_{xyzt}$; this latter is known to satisfy the the
so--called \textit{real special geometry constraints} (see \textit{e.g.}
Eq.~(2.12) of \cite{Cerchiai:2010xv}).} (general results on supersymmetry
iterations at the third and fourth order are given in Apps. \ref{3rd-Order}
and \ref{4th-Order}, respectively) :
\begin{align}
\left( \delta ^{\left( 2\right) }e_{\mu }^{a}\right) =& \frac{1}{2}\bar{%
\epsilon}\gamma ^{a}\left( \delta ^{\left( 1\right) }\psi _{\mu }\right) \,,
\\
\left( \delta ^{\left( 2\right) }\psi _{\mu }^{i}\right) =& \left( \delta
^{\left( 1\right) }{\mathcal{D}}_{\mu }\right) \epsilon ^{i}-\frac{1}{6}%
\epsilon _{j}\bar{\lambda}^{ix}\gamma _{\mu }\left( \delta ^{\left( 1\right)
}\lambda _{x}^{j}\right) +\frac{1}{12}\gamma _{\mu \nu }\epsilon _{j}\bar{%
\lambda}^{ix}\gamma ^{\nu }\left( \delta ^{\left( 1\right) }\lambda
_{x}^{j}\right) +  \notag \\
& -\frac{1}{48}\gamma _{\mu \nu \rho }\epsilon _{j}\bar{\lambda}^{ix}\gamma
^{\nu \rho }\left( \delta ^{\left( 1\right) }\lambda _{x}^{j}\right) +\frac{1%
}{12}\gamma ^{\nu }\bar{\lambda}^{ix}\gamma _{\mu \nu }\left( \delta
^{\left( 1\right) }\lambda _{x}^{j}\right) +  \notag \\
& -\frac{1}{6}\epsilon _{j}\bar{\lambda}^{ix}\gamma _{a}\lambda
_{x}^{j}\left( \delta ^{\left( 1\right) }e_{\mu }^{a}\right) -\frac{1}{6}%
\epsilon _{j}\left( \delta ^{\left( 1\right) }\bar{\lambda}^{ix}\right)
\gamma _{\mu }\lambda _{x}^{j}+  \notag \\
& +\frac{1}{12}\gamma _{ab}\epsilon _{j}\bar{\lambda}^{ix}\gamma ^{b}\lambda
_{x}^{j}\left( \delta ^{\left( 1\right) }e_{\mu }^{a}\right) +\frac{1}{12}%
\gamma _{\mu \nu }\epsilon _{j}\left( \delta ^{\left( 1\right) }\bar{\lambda}%
^{ix}\right) \gamma ^{\nu }\lambda _{x}^{j}+  \notag \\
& -\frac{1}{48}\gamma _{abc}\epsilon _{j}\bar{\lambda}^{ix}\gamma
^{bc}\lambda _{x}^{j}\left( \delta ^{\left( 1\right) }e_{\mu }^{a}\right) -%
\frac{1}{48}\gamma _{\mu \nu \rho }\epsilon _{j}\left( \delta ^{\left(
1\right) }\bar{\lambda}^{ix}\right) \gamma ^{\nu \rho }\lambda _{x}^{j}+
\notag \\
& +\frac{1}{12}\gamma ^{b}\epsilon _{j}\bar{\lambda}^{ix}\gamma _{ab}\lambda
_{x}^{j}\left( \delta ^{\left( 1\right) }e_{\mu }^{a}\right) +\frac{1}{12}%
\gamma ^{\nu }\epsilon _{j}\left( \delta ^{\left( 1\right) }\bar{\lambda}%
^{ix}\right) \gamma _{\mu \nu }\lambda _{x}^{j}+  \notag \\
& +\frac{i}{4\sqrt{6}}h_{I}\tilde{F}_{\nu \rho }^{I}\left[ \left( \delta
^{\left( 1\right) }e_{\mu }^{a}\right) e_{b}^{\nu }e_{c}^{\rho }+e_{\mu
}^{a}\left( \delta ^{\left( 1\right) }e_{b}^{\nu }\right) e_{c}^{\rho
}+e_{\mu }^{a}e_{b}^{\nu }\left( \delta ^{\left( 1\right) }e_{c}^{\rho
}\right) \right] \left( \gamma _{a}^{\phantom{a}bc}-4\delta _{a}^{b}\gamma
^{c}\right) \epsilon ^{i}+  \notag \\
& +\frac{i}{12}h_{Iz}\left( \delta ^{\left( 1\right) }\phi ^{z}\right)
\tilde{F}_{\nu \rho }^{I}\left( \gamma _{\mu }^{\phantom{\m}\nu \rho
}-4\delta _{\mu }^{\nu }\gamma ^{\rho }\right) \epsilon ^{i}  \notag \\
& +\frac{i}{4\sqrt{6}}h_{I}\left( \delta ^{\left( 1\right) }\tilde{F}_{\nu
\rho }^{I}\right) \left( \gamma _{\mu }^{\phantom{\m}\nu \rho }-4\delta
_{\mu }^{\nu }\gamma ^{\rho }\right) \epsilon ^{i}\,, \\
\left( \delta ^{\left( 2\right) }\phi ^{x}\right) =& \frac{i}{2}\bar{\epsilon%
}\left( \delta ^{\left( 1\right) }\lambda ^{x}\right) \,, \\
\left( \delta ^{\left( 2\right) }A_{\mu }^{I}\right) =& -\frac{1}{2}\bar{%
\epsilon}\gamma _{\mu }\left( \delta ^{\left( 1\right) }\lambda ^{x}\right)
h_{x}^{I}-\frac{1}{2}\left( \delta ^{\left( 1\right) }e_{\mu }^{a}\right)
\bar{\epsilon}\gamma _{a}\lambda ^{x}h_{x}^{I}+  \notag \\
& -\frac{i}{2}\sqrt{\frac{3}{2}}\bar{\epsilon}h^{I}\left( \delta ^{\left(
1\right) }\psi _{\mu }\right) +\frac{i}{2}h_{x}^{I}\left( \delta ^{\left(
1\right) }\phi ^{x}\right) \bar{\epsilon}\psi _{\mu }+  \notag \\
& -\frac{1}{2}\bar{\epsilon}\gamma _{\mu }\lambda ^{x}\nabla
_{y}h_{x}^{I}\left( \delta ^{\left( 1\right) }\phi ^{y}\right) \,, \\
\left( \delta ^{\left( 2\right) }\lambda ^{ix}\right) =& -\frac{i}{2}\left(
\delta ^{\left( 1\right) }e_{a}^{\mu }\right) \gamma ^{a}\widehat{\mathcal{D}%
}_{\mu }\phi ^{x}\epsilon ^{i}-\frac{i}{2}\gamma ^{\mu }\left( \delta
^{\left( 1\right) }\widehat{\mathcal{D}}_{\mu }\phi ^{x}\right) \epsilon
^{i}-\frac{1}{4\sqrt{6}}T^{xyz}\gamma ^{\mu }\epsilon _{j}\bar{\lambda}%
_{y}^{i}\gamma _{\mu }\left( \delta ^{\left( 1\right) }\lambda
_{z}^{j}\right) +  \notag \\
& +\frac{1}{4}\sqrt{\frac{3}{2}}T^{xyz}\epsilon _{j}\bar{\lambda}%
_{y}^{i}\left( \delta ^{\left( 1\right) }\lambda _{z}^{j}\right) -\frac{1}{8%
\sqrt{6}}T^{xyz}\gamma ^{\mu \nu }\epsilon _{j}\bar{\lambda}_{y}^{i}\gamma
_{\mu \nu }\left( \delta ^{\left( 1\right) }\lambda _{z}^{j}\right) +  \notag
\\
& -\left( \delta ^{\left( 1\right) }\phi ^{y}\right) \Gamma _{yz}^{x}\left(
\delta ^{\left( 1\right) }\lambda ^{zi}\right) -\left( \delta ^{\left(
2\right) }\phi ^{y}\right) \Gamma _{yz}^{x}\lambda ^{zi}+\frac{1}{4}\sqrt{%
\frac{3}{2}}T^{xyz}\epsilon _{j}\left( \delta ^{\left( 1\right) }\bar{\lambda%
}_{y}^{i}\right) \lambda _{z}^{j}+  \notag \\
& -\frac{1}{8\sqrt{6}}T^{xyz}\gamma ^{\mu \nu }\epsilon _{j}\left( \delta
^{\left( 1\right) }\bar{\lambda}_{y}^{i}\right) \gamma _{\mu \nu }\lambda
_{z}^{j}-\frac{1}{4\sqrt{6}}T^{xyz}\gamma ^{\mu }\epsilon _{j}\left( \delta
^{\left( 1\right) }\bar{\lambda}_{y}^{i}\right) \gamma _{\mu }\lambda
_{z}^{j}+  \notag \\
& +\frac{1}{4\sqrt{6}}\nabla _{t}T^{xyz}\left( \delta ^{\left( 1\right)
}\phi ^{t}\right) \left[ 3\epsilon _{j}\bar{\lambda}_{y}^{i}\lambda
_{z}^{j}-\gamma ^{\mu }\epsilon _{j}\bar{\lambda}_{y}^{i}{\gamma }_{\mu
}\lambda _{z}^{j}-{\textstyle\frac{1}{2}}\gamma ^{\mu \nu }\epsilon _{j}\bar{%
\lambda}_{y}^{i}{\gamma }_{\mu \nu }\lambda _{z}^{j}\right] +  \notag \\
& -\left( \delta ^{\left( 1\right) }\phi ^{y}\right) \nabla _{t}\Gamma
_{yz}^{x}\left( \delta ^{\left( 1\right) }\phi ^{t}\right) \lambda ^{zi}+%
\frac{1}{4}\gamma \cdot \tilde{F}^{I}\nabla _{t}h_{I}^{x}\left( \delta
^{\left( 1\right) }\phi ^{t}\right) \epsilon ^{i}+\frac{1}{4}\gamma \cdot
\left( \delta ^{\left( 1\right) }\tilde{F}^{I}\right) h_{I}^{x}\epsilon ^{i}+
\notag \\
& +\frac{1}{2}\left( \delta ^{\left( 1\right) }e_{a}^{\mu }\right)
e_{b}^{\nu }+e_{a}^{\mu }\left( \delta ^{\left( 1\right) }e_{b}^{\nu
}\right) \gamma ^{ab}\tilde{F}_{\mu \nu }^{I}h_{I}^{x}\epsilon ^{i}\ .
\end{align}%
with

\begin{align}
\left(\delta^{\left(1\right)}\tilde{F}^I_{\mu\nu}\right) = &
\left(\delta^{\left(1\right)}\mathcal{F}^I_{\mu\nu}\right) +
\left(\delta^{\left(1\right)}\bar{\psi}_{\left[\mu\right.}\right)
\gamma_{\left.\nu\right]} \lambda^x h^I_x + \bar{\psi}_{\left[\mu\right.}
\gamma_{\left.\nu\right]} \left(\delta^{\left(1\right)}\lambda^x\right)
h^I_x +  \notag \\
& + \bar{\psi}_{\left[\mu\right.} \gamma_{\left.\nu\right]} \lambda^x
\nabla_y h^I_x \left(\delta^{\left(1\right)}\phi^y\right) + \frac{i \sqrt{6}%
}{4 } \left(\delta^{\left(1\right)}\bar{\psi}_{\mu}\right) \psi_{\nu} h^I +
\notag \\
& + \frac{i \sqrt{6}}{4 } \bar{\psi}_{\mu}
\left(\delta^{\left(1\right)}\psi_{\nu}\right) h^I + \bar{\psi}_{\left[%
\mu\right.} \left(\delta^{\left(1\right)}e^a_{\left.\nu\right]}\right)
\gamma_a \lambda^x h^I_x +  \notag \\
& -\frac{i}{2} \bar{\psi}_{\mu} \psi_{\nu} h^I_x
\left(\delta^{\left(1\right)}\phi^x\right) \,, \\
\left(\delta^{\left(1\right)}{\mathcal{D}}_{\mu}\right) = & \frac{1}{4}
\left(\delta^{\left(1\right)}\omega^{ab}_\mu\right) \gamma_{ab} \,, \\
\left(\delta^{\left(1\right)}\omega^{ab}_\mu\right) = & \frac{1}{2}
\left(\delta^{\left(1\right)}e_{c\mu}\right) \left[ \Omega^{abc} -
\Omega^{bca} - \Omega^{cab} \right] +  \notag \\
& + \frac{1}{2} e_{c\mu} \left[ \left(\delta^{\left(1\right)}\Omega^{abc}%
\right) - \left(\delta^{\left(1\right)}\Omega^{bca}\right) -
\left(\delta^{\left(1\right)}\Omega^{cab}\right)\right] +
\left(\delta^{\left(1\right)}K^{a \phantom{\mu} b}_{\phantom{a}\mu}\right)
\,, \\
\left(\delta^{\left(1\right)}\Omega^{abc}\right)= &\left[ \left(\delta^{%
\left(1\right)}e^{\mu a}\right) e^{\nu b} +e^{\mu a}
\left(\delta^{\left(1\right)}e^{\nu b}\right) \right] \left( \partial_{\mu}
e^{c}_{\nu} - \partial_{\nu} e^{c}_{\mu} \right) +  \notag \\
& + e^{\mu a} e^{\nu b} \left[ \partial_{\mu}
\left(\delta^{\left(1\right)}e^{c}_{\nu}\right) - \partial_{\nu}
\left(\delta^{\left(1\right)}e^{c}_{\mu}\right) \right] \,, \\
\left(\delta^{\left(1\right)}K^{a \phantom{\mu} b}_{\phantom{a}\mu}\right) =
& \frac{1}{2} \left[\left(\delta^{\left(1\right)}\bar{\psi}_{ \rho}\right)
e^{\rho \left[a\right.} \gamma^{\left. b \right]} \psi_{\mu} + \bar{\psi}_{
\rho} \left(\delta^{\left(1\right)}e^{\rho \left[a\right.}\right)
\gamma^{\left. b \right]} \psi_{\mu} + \right.  \notag \\
& + \bar{\psi}^{\left[a\right.} \gamma^{\left. b \right]}
\left(\delta^{\left(1\right)}\psi_{\mu}\right) + \frac{1}{2}
\left(\delta^{\left(1\right)}\bar{\psi}_{ \rho}\right) e^{\rho a}
\gamma_{\mu} \psi^{ b} +  \notag \\
& + \frac{1}{2} \bar{\psi}_{ \rho} \left(\delta^{\left(1\right)}e^{\rho
a}\right) \gamma_{\mu} \psi^{ b} + \frac{1}{2} \bar{\psi}^a \gamma_{a}
\left(\delta^{\left(1\right)}e^a_{\mu}\right) \psi^{ b} +  \notag \\
& + \left. \frac{1}{2} \bar{\psi}^a \gamma_{\mu} \psi_{\rho}
\left(\delta^{\left(1\right)}e^{\rho b}\right) + \frac{1}{2} \bar{\psi}^a
\gamma_\mu \left(\delta^{\left(1\right)}\psi_\rho\right) e^{\rho b} \right]
\,, \\
\left(\delta^{\left(1\right)}\widehat{{\mathcal{D}}}_\mu \phi^x\right) = &
\partial_\mu \left(\delta^{\left(1\right)}\phi^x\right) - \frac{i}{2}
\left(\delta^{\left(1\right)}\bar{\psi}_\mu\right)\lambda^x - \frac{i}{2}
\bar{\psi}_\mu \left(\delta^{\left(1\right)}\lambda^x\right) \ .
\end{align}

\section{\label{Purely-Bosonic}Evaluation on Purely Bosonic Background}

Next, we proceed to evaluate the \textit{fermionic wigging} on a \textit{%
purely bosonic} background (characterized by setting $\psi =\lambda =0$
identically, and denoted by $|_{\mathrm{bg}}$ throughout). This results in a
dramatic simplification of previous formul\ae ; in particular, all covariant
quantities, such as the $\widetilde{E}$--tensor \cite{Cerchiai:2010xv},
characterizing the real special geometry of the scalar manifold (\textit{cfr.%
} Apps. \ref{3rd-Order} and \ref{4th-Order}), do not occur anymore after
evaluation on such a background.

\subsection{\label{1st-Order-PB}First Order}

At the first order, the non-zero supersymmetry variations are:
\begin{subequations}
\begin{align}
\left. \left( \delta ^{\left( 1\right) }\psi^i _{\mu }\right) \right\vert _{%
\mathrm{bg}}=& D_{\mu }\left( \hat{\omega}\right) \epsilon ^{i}+\frac{i}{4%
\sqrt{6}}h_{I}\mathcal{F}_{\nu \rho }^{I}\left( \gamma _{\mu }^{\phantom{\m}%
\nu \rho }-4\delta _{\mu }^{\nu }\gamma ^{\rho }\right) \epsilon ^{i}\,, \\
\left. \left( \delta ^{\left( 1\right) }\lambda ^{xi}\right) \right\vert _{%
\mathrm{bg}}=& -\frac{i}{2}\slashed{\partial}\phi ^{x}\epsilon ^{i}+\frac{1}{%
4}\gamma \cdot \mathcal{F}^{I}h_{I}^{x}\epsilon ^{i}\ .
\end{align}%
Moreover, the supercovariant field strength collapses to the ordinary field
strength and the covariant derivative on $\phi ^{x}$ reduces to an ordinary
(flat) derivative.

\subsection{Second Order}

\end{subequations}
\begin{subequations}
\begin{align}
\left. \left( \delta ^{\left( 2\right) }e_{\mu }^{a}\right) \right\vert _{%
\mathrm{bg}}=& \frac{1}{2}\bar{\epsilon}\gamma ^{a}\left. \left( \delta
^{\left( 1\right) }\psi _{\mu }\right) \right\vert _{\mathrm{bg}}\ , \\
\left. \left( \delta ^{\left( 2\right) }\phi ^{x}\right) \right\vert _{%
\mathrm{bg}}=& \frac{i}{2}\bar{\epsilon}\left. \left( \delta ^{\left(
1\right) }\lambda ^{x}\right) \right\vert _{\mathrm{bg}}\ , \\
\left. \left( \delta ^{\left( 2\right) }A_{\mu }^{I}\right) \right\vert _{%
\mathrm{bg}}=& -\frac{i}{2}\sqrt{\frac{3}{2}}\bar{\epsilon}h^{I}\left.
\left( \delta ^{\left( 1\right) }\psi _{\mu }\right) \right\vert _{\mathrm{bg%
}}-\frac{1}{2}\bar{\epsilon}\gamma _{\mu }\left. \left( \delta ^{\left(
1\right) }\lambda ^{x}\right) \right\vert _{\mathrm{bg}}h_{x}^{I}\ .
\end{align}%
The supercovariant field strength, the covariant derivative on $\phi ^{x}$
and the variation of the spin connection $\omega _{\mu }^{ab}$ all collapse
to zero.

\subsection{Third Order}

At the third order, one obtains the following results :
\end{subequations}
\begin{subequations}
\begin{align}
\left. \left( \delta ^{\left( 3\right) }\psi^i_{\mu }\right) \right\vert _{%
\mathrm{bg}}=& \left. \left( \delta ^{\left( 2\right) }{\mathcal{D}}_{\mu
}\right) \right\vert _{\mathrm{bg}}\epsilon^i -\frac{1}{3}\epsilon
_{j}\left. \left( \delta ^{\left( 1\right) }\bar{\lambda}^{ix}\right)
\right\vert _{\mathrm{bg}}\gamma _{\mu }\left. \left( \delta ^{\left(
1\right) }\lambda _{x}^{j}\right) \right\vert _{\mathrm{bg}}+  \notag \\
& +\frac{1}{6}\gamma _{\mu \nu }\left. \left( \delta ^{\left( 1\right) }\bar{%
\lambda}^{ix}\right) \right\vert _{\mathrm{bg}}\gamma ^{\nu }\left. \left(
\delta ^{\left( 1\right) }\lambda _{x}^{j}\right) \right\vert _{\mathrm{bg}}+
\notag \\
& -\frac{1}{24}\gamma _{\mu \nu \rho }\epsilon _{j}\left. \left( \delta
^{\left( 1\right) }\bar{\lambda}^{ix}\right) \right\vert _{\mathrm{bg}%
}\gamma ^{\nu \rho }\left. \left( \delta ^{\left( 1\right) }\lambda
_{x}^{j}\right) \right\vert _{\mathrm{bg}}+  \notag \\
& +\frac{1}{6}\gamma _{\mu \nu }\epsilon _{j}\left. \left( \delta ^{\left(
1\right) }\bar{\lambda}^{ix}\right) \right\vert _{\mathrm{bg}}\gamma ^{\nu
}\left. \left( \delta ^{\left( 1\right) }\lambda _{x}^{j}\right) \right\vert
_{\mathrm{bg}}+  \notag \\
& +\frac{i}{4\sqrt{6}}h_{I}\mathcal{F}_{\nu\rho}^{I}\left[ \left. \left(
\delta ^{\left( 2\right) }e_{\mu }^{a}\right) \right\vert _{\mathrm{bg}%
}e_{b}^{\nu }e_{c}^{\rho }+e_{\mu }^{a}\left. \left( \delta ^{\left(
2\right) }e_{b}^{\nu }\right) \right\vert _{\mathrm{bg}}e_{c}^{\rho }+e_{\mu
}^{a}e_{b}^{\nu }\left. \left( \delta ^{\left( 2\right) }e_{c}^{\rho
}\right) \right\vert _{\mathrm{bg}}\right] \times  \notag \\
& \times \left( \gamma _{a}^{\phantom{a}bc}-4\delta _{a}^{b}\gamma
^{c}\right) \epsilon ^{i}+\frac{i}{12}h_{Iz}\left. \left( \delta ^{\left(
2\right) }\phi ^{z}\right) \right\vert _{\mathrm{bg}}\mathcal{F}_{\nu \rho
}^{I}\left( \gamma _{\mu }^{\phantom{\m}\nu \rho }-4\delta _{\mu }^{\nu
}\gamma ^{\rho }\right) \epsilon ^{i}+  \notag \\
& +\frac{i}{4\sqrt{6}}h_{I}\left. \left( \delta ^{\left( 2\right) }\tilde{F}%
_{\nu \rho }^{I}\right) \right\vert _{\mathrm{bg}}\left( \gamma _{\mu }^{%
\phantom{\m}\nu \rho }-4\delta _{\mu }^{\nu }\gamma ^{\rho }\right) \epsilon
^{i} \,, \\
\left. \left( \delta ^{\left( 3\right) }\lambda ^{ix}\right) \right\vert _{%
\mathrm{bg}}=& -\frac{i}{2}\left. \left( \delta ^{\left( 2\right)
}e_{a}^{\mu }\right) \right\vert _{\mathrm{bg}}\gamma ^{a}\partial _{\mu
}\phi ^{x}\epsilon ^{i}-\frac{i}{2}\gamma ^{\mu }\left. \left( \delta
^{\left( 2\right) }\widehat{\mathcal{D}}_{\mu }\phi ^{x}\right) \right\vert
_{\mathrm{bg}}\epsilon ^{i}-2\left. \left( \delta ^{\left( 2\right) }\phi
^{y}\right) \right\vert _{\mathrm{bg}}\Gamma _{yz}^{x}\left. \left( \delta
^{\left( 1\right) }\lambda ^{zi}\right) \right\vert _{\mathrm{bg}}+  \notag
\\
& +\frac{1}{2}\sqrt{\frac{3}{2}}T^{xyz}\epsilon _{j}\left. \left( \delta
^{\left( 1\right) }\bar{\lambda}_{y}^{i}\right) \right\vert _{\mathrm{bg}%
}\left. \left( \delta ^{\left( 1\right) }\lambda _{z}^{j}\right) \right\vert
_{\mathrm{bg}}+  \notag \\
& -\frac{1}{4\sqrt{6}}T^{xyz}\gamma ^{\mu \nu }\epsilon _{j}\left. \left(
\delta ^{\left( 1\right) }\bar{\lambda}_{y}^{i}\right) \right\vert _{\mathrm{%
bg}}\gamma _{\mu \nu }\left. \left( \delta ^{\left( 1\right) }\lambda
_{z}^{j}\right) \right\vert _{\mathrm{bg}}+  \notag \\
& -\frac{1}{2\sqrt{6}}T^{xyz}\gamma ^{\mu }\epsilon _{j}\left. \left( \delta
^{\left( 1\right) }\bar{\lambda}_{y}^{i}\right) \right\vert _{\mathrm{bg}%
}\gamma _{\mu }\left. \left( \delta ^{\left( 1\right) }\lambda
_{z}^{j}\right) \right\vert _{\mathrm{bg}}+  \notag \\
& +\frac{1}{4}\gamma \cdot \mathcal{F}^{I}\nabla _{t}h_{I}^{x}\left. \left(
\delta ^{\left( 2\right) }\phi ^{t}\right) \right\vert _{\mathrm{bg}%
}\epsilon ^{i}+\frac{1}{4}\gamma \cdot \left. \left( \delta ^{\left(
2\right) }\tilde{F}^{I}\right) \right\vert _{\mathrm{bg}}h_{I}^{x}\epsilon
^{i}+  \notag \\
& +\frac{1}{4}\gamma ^{ab}\left[ \left. \left( \delta ^{\left( 2\right)
}e_{a}^{\mu }\right) \right\vert _{\mathrm{bg}}e_{b}^{\nu }+e_{a}^{\mu
}\left. \left( \delta ^{\left( 2\right) }e_{b}^{\nu }\right) \right\vert _{%
\mathrm{bg}}\right] \mathcal{F}_{\mu \nu }^{I}h_{I}^{x}\epsilon ^{i}\ .
\end{align}%
For the supercovariant field strength, the covariant derivative on $\phi
^{x} $, and the spin connection $\omega _{\mu }^{ab}$, it holds that:
\end{subequations}
\begin{subequations}
\begin{align}
\left. \left( \delta ^{\left( 2\right) }\tilde{F}_{\mu \nu }^{I}\right)
\right\vert _{\mathrm{bg}}=& 2\partial _{\left[ \mu \right. }\left. \left(
\delta ^{\left( 2\right) }A_{\left. \nu \right] }^{I}\right) \right\vert _{%
\mathrm{bg}}+2\left. \left( \delta ^{\left( 1\right) }\bar{\psi}_{\left[ \mu
\right. }\right) \right\vert _{\mathrm{bg}}\gamma _{\left. \nu \right]
}\left. \left( \delta ^{\left( 1\right) }\lambda ^{x}\right) \right\vert _{%
\mathrm{bg}}h_{x}^{I}+  \notag \\
& + i \sqrt{\frac{3}{2}}\left. \left( \delta ^{\left( 1\right) }\bar{\psi}%
_{\nu }\right) \right\vert _{\mathrm{bg}}\left. \left( \delta ^{\left(
1\right) }\psi _{\mu }\right) \right\vert _{\mathrm{bg}}h^{I} \,, \\
\left. \left( \delta ^{\left( 2\right) }\widehat{{\mathcal{D}}}_{\mu }\phi
^{x}\right) \right\vert _{\mathrm{bg}}=& \partial _{\mu }\left. \left(
\delta ^{\left( 2\right) }\phi ^{x}\right) \right\vert _{\mathrm{bg}%
}-i\left. \left( \delta ^{\left( 1\right) }\bar{\psi}_{\mu }\right)
\right\vert _{\mathrm{bg}}\left. \left( \delta ^{\left( 1\right) }\lambda
^{x}\right) \right\vert _{\mathrm{bg}} \,, \\
\left. \left( \delta ^{\left( 2\right) }\omega _{\mu }^{ab}\right)
\right\vert _{\mathrm{bg}}=& \frac{1}{2}\left. \left( \delta ^{\left(
2\right) }e_{c\mu }\right) \right\vert _{\mathrm{bg}}\left( \Omega
^{abc}-\Omega ^{bca}-\Omega ^{cab}\right) +  \notag \\
& +\frac{1}{2}\left[ \left. \left( \delta ^{\left( 2\right) }\Omega
^{abc}\right) \right\vert _{\mathrm{bg}}-\left. \left( \delta ^{\left(
2\right) }\Omega ^{bca}\right) \right\vert _{\mathrm{bg}}-\left. \left(
\delta ^{\left( 2\right) }\Omega ^{cab}\right) \right\vert _{\mathrm{bg}}%
\right] +\left. \left( \delta ^{\left( 2\right) }K_{\phantom{a}\mu }^{a%
\phantom{\mu}b}\right) \right\vert _{\mathrm{bg}} \,, \\
\left. \left( \delta ^{\left( 2\right) }\Omega ^{abc}\right) \right\vert _{%
\mathrm{bg}}=& \left[ \left. \left( \delta ^{\left( 2\right) }e^{\mu
a}\right) \right\vert _{\mathrm{bg}}e^{\nu b}+e^{\mu a}\left. \left( \delta
^{\left( 2\right) }e^{\nu b}\right) \right\vert _{\mathrm{bg}}\right] \left(
\partial _{\mu }e_{\nu }^{c}-\partial _{\nu }e_{\mu }^{c}\right) +  \notag \\
& +e^{\mu a}e^{\nu b}\left[ \partial _{\mu }\left. \left( \delta ^{\left(
2\right) }e_{\nu }^{c}\right) \right\vert _{\mathrm{bg}}-\partial _{\nu
}\left. \left( \delta ^{\left( 2\right) }e_{\mu }^{c}\right) \right\vert _{%
\mathrm{bg}}\right] \,, \\
\left. \left( \delta ^{\left( 2\right) }K_{\phantom{a}\mu }^{a\phantom{\mu}%
b}\right) \right\vert _{\mathrm{bg}}=& \left. \left( \delta ^{\left(
1\right) }\bar{\psi}_{\rho }\right) \right\vert _{\mathrm{bg}}e^{\rho \left[
a\right. }\gamma ^{\left. b\right] }\left. \left( \delta ^{\left( 1\right)
}\psi _{\mu }\right) \right\vert _{\mathrm{bg}}+  \notag \\
& +\frac{1}{2}\left. \left( \delta ^{\left( 1\right) }\bar{\psi}_{\rho
}\right) \right\vert _{\mathrm{bg}}\gamma _{\mu }\left. \left( \delta
^{\left( 1\right) }\psi _{\nu }\right) \right\vert _{\mathrm{bg}}e^{\rho
a}e^{\nu b}\ .
\end{align}

\subsection{Fourth Order}

Finally, at the fourth order, one achieves the following expressions :
\end{subequations}
\begin{subequations}
\begin{align}
\left. \left( \delta ^{\left( 4\right) }e_{\mu }^{a}\right) \right\vert _{%
\mathrm{bg}}=& \frac{1}{2}\bar{\epsilon}\gamma ^{a}\left. \left( \delta
^{\left( 3\right) }\psi _{\mu }\right) \right\vert _{\mathrm{bg}}\,, \\
\left. \left( \delta ^{\left( 4\right) }\phi ^{x}\right) \right\vert _{%
\mathrm{bg}}=& \frac{i}{2}\bar{\epsilon}\left. \left( \delta ^{\left(
3\right) }\lambda ^{x}\right) \right\vert _{\mathrm{bg}}\,, \\
\left. \left( \delta ^{\left( 4\right) }A_{\mu }^{I}\right) \right\vert _{%
\mathrm{bg}}=& -\frac{1}{2}\bar{\epsilon}\gamma _{\mu }\left. \left( \delta
^{\left( 3\right) }\lambda ^{x}\right) \right\vert _{\mathrm{bg}}h_{x}^{I}+
\notag \\
& -\frac{i}{2}\sqrt{\frac{3}{2}}\bar{\epsilon}h^{I}\left. \left( \delta
^{\left( 3\right) }\psi _{\mu }\right) \right\vert _{\mathrm{bg}}+  \notag \\
& +\frac{3i}{2}h_{x}^{I}\left. \left( \delta ^{\left( 2\right) }\phi
^{x}\right) \right\vert _{\mathrm{bg}}\bar{\epsilon}\left. \left( \delta
^{\left( 1\right) }\psi _{\mu }\right) \right\vert _{\mathrm{bg}}+  \notag \\
& -\frac{3}{2}\left. \left( \delta ^{\left( 2\right) }e_{\mu }^{a}\right)
\right\vert _{\mathrm{bg}}\bar{\epsilon}\gamma _{a}\left. \left( \delta
^{\left( 1\right) }\lambda ^{x}\right) \right\vert _{\mathrm{bg}}h_{x}^{I}+
\notag \\
& -\frac{3}{2}\bar{\epsilon}\gamma _{\mu }\left. \left( \delta ^{\left(
1\right) }\lambda ^{x}\right) \right\vert _{\mathrm{bg}}\nabla
_{y}h_{x}^{I}\left. \left( \delta ^{\left( 2\right) }\phi ^{y}\right)
\right\vert _{\mathrm{bg}}\ .
\end{align}%
Again, the supercovariant field strength, the covariant derivative on $\phi
^{x}$ and the spin connection $\omega _{\mu }^{ab}$ all vanish.

\section{\label{Wig-BPS-D=5}Wigging of BPS Extremal Black Hole}

Following the treatment of the $D=5$ \textit{attractor mechanism} given in
\cite{Chamseddine:1996pi,Chamseddine:1999qs} and \cite{Kraus:2005gh}, we
consider the $1/2$--BPS near--horizon conditions for extremal electric black
hole (with near-horizon geometry $AdS_{2}\times S^{3}$):

\end{subequations}
\begin{align}
& \partial _{\mu }h^{I}=0\Longrightarrow \partial _{\mu }\phi ^{x}=0\ ,
\notag \\
& h_{Ix}F_{\mu \nu }^{I}=0\ ,  \label{extrBHcond}
\end{align}%
and we evaluate the results for purely bosonic background (computed in the
previous section) onto such conditions (denoted by $|_{BPS}$, and always
understood on the r.h.s. of equations, throughout the following treatment).

\subsection{\label{1st-Order-BPS}First Order}

At the first order, the gravitino variation is non--zero, while the gaugino
variation vanishes :
\begin{subequations}
\begin{align}
\left. \left( \delta ^{\left( 1\right) }\psi^i_{\mu }\right) \right\vert _{%
\mathrm{BPS}}=& D_{\mu }\left( \hat{\omega}\right) \epsilon ^{i}+\frac{i}{4%
\sqrt{6}}h_{I}\mathcal{F}_{\nu \rho }^{I}\left( \gamma _{\mu }^{\phantom{\m}%
\nu \rho }-4\delta _{\rho }^{\nu }\gamma ^{\rho }\right) \epsilon ^{i}\neq 0
\,, \\
\left. \left( \delta ^{\left( 1\right) }\lambda ^{xi}\right) \right\vert _{%
\mathrm{BPS}}=&~ 0\ .
\end{align}

\subsection{\label{52}Second Order}

At the second order, one obtains :
\end{subequations}
\begin{subequations}
\begin{align}
\left. \left( \delta ^{\left( 2\right) }e_{\mu }^{a}\right) \right\vert _{%
\mathrm{BPS}}=& \frac{1}{2}\bar{\epsilon}\gamma ^{a}\left. \left( \delta
^{\left( 1\right) }\psi _{\mu }\right) \right\vert _{\mathrm{BPS}}\neq 0 \,,
\\
\left. \left( \delta ^{\left( 2\right) }\phi ^{x}\right) \right\vert _{%
\mathrm{BPS}}=& ~0 \,, \\
\left. \left( \delta ^{\left( 2\right) }A_{\mu }^{I}\right) \right\vert _{%
\mathrm{BPS}}=& ~0 \ .
\end{align}

\subsection{Third Order}

At the third order, it holds that :
\end{subequations}
\begin{subequations}
\begin{align}
\left. \left( \delta ^{\left( 3\right) }\psi _{\mu }\right) \right\vert _{%
\mathrm{BPS}}=& \left. \left( \delta ^{\left( 2\right) }{\mathcal{D}}_{\mu
}\right) \right\vert _{\mathrm{BPS}}\epsilon +  \notag \\
& +\frac{i}{4\sqrt{6}}h_{I}\mathcal{F}_{bc}^{I}\left[ \left. \left( \delta
^{\left( 2\right) }e_{\mu }^{a}\right) \right\vert _{\mathrm{BPS}}e_{b}^{\nu
}e_{c}^{\rho }+e_{\mu }^{a}\left. \left( \delta ^{\left( 2\right)
}e_{b}^{\nu }\right) \right\vert _{\mathrm{BPS}}e_{c}^{\rho }+e_{\mu
}^{a}e_{b}^{\nu }\left. \left( \delta ^{\left( 2\right) }e_{c}^{\rho
}\right) \right\vert _{\mathrm{BPS}}\right] \times  \notag \\
& \times \left( \gamma _{a}^{\phantom{a}bc}-4\delta _{a}^{b}\gamma
^{c}\right) \epsilon ^{i}+\frac{i}{4\sqrt{6}}h_{I}\left. \left( \delta
^{\left( 2\right) }\tilde{F}_{\nu \rho }^{I}\right) \right\vert _{\mathrm{BPS%
}}\left( \gamma _{\mu }^{\phantom{\m}\nu \rho }-4\delta _{\mu }^{\nu }\gamma
^{\rho }\right) \epsilon ^{i}\,,
\end{align}%
and
\begin{equation}
\left. \left( \delta ^{\left( 3\right) }\lambda ^{ix}\right) \right\vert _{%
\mathrm{BPS}}=0\ .
\end{equation}%
Concerning the supercovariant field strength, the covariant derivative on $%
\phi ^{x}$ and the spin connection $\omega _{\mu }^{ab}$, the following
expressions hold:
\end{subequations}
\begin{subequations}
\begin{align}
\left. \left( \delta ^{\left( 2\right) }\tilde{F}_{\mu \nu }^{I}\right)
\right\vert _{\mathrm{BPS}}=& 2\partial _{\left[ \mu \right. }\left. \left(
\delta ^{\left( 2\right) }A_{\left. \nu \right] }^{I}\right) \right\vert _{%
\mathrm{BPS}}+i\sqrt{\frac{3}{2}}\left. \left( \delta ^{\left( 1\right) }%
\bar{\psi}_{\nu }\right) \right\vert _{\mathrm{BPS}}\left. \left( \delta
^{\left( 1\right) }\psi _{\mu }\right) \right\vert _{\mathrm{BPS}}h^{I}\,, \\
\left. \left( \delta ^{\left( 2\right) }\widehat{{\mathcal{D}}}_{\mu }\phi
^{x}\right) \right\vert _{\mathrm{BPS}}=& \partial _{\mu }\left. \left(
\delta ^{\left( 2\right) }\phi ^{x}\right) \right\vert _{\mathrm{BPS}%
}-i\left. \left( \delta ^{\left( 1\right) }\bar{\psi}_{\mu }\right)
\right\vert _{\mathrm{BPS}}\left. \left( \delta ^{\left( 1\right) }\lambda
^{x}\right) \right\vert _{\mathrm{BPS}}=0\,, \\
\left. \left( \delta ^{\left( 2\right) }\omega _{\mu }^{ab}\right)
\right\vert _{\mathrm{BPS}}=& \frac{1}{2}\left. \left( \delta ^{\left(
2\right) }e_{c\mu }\right) \right\vert _{\mathrm{BPS}}\left( \Omega
^{abc}-\Omega ^{bca}-\Omega ^{cab}\right) +  \notag \\
& +\frac{1}{2}\left[ \left. \left( \delta ^{\left( 2\right) }\Omega
^{abc}\right) \right\vert _{\mathrm{BPS}}-\left. \left( \delta ^{\left(
2\right) }\Omega ^{bca}\right) \right\vert _{\mathrm{BPS}}-\left. \left(
\delta ^{\left( 2\right) }\Omega ^{cab}\right) \right\vert _{\mathrm{BPS}}%
\right] +\left. \left( \delta ^{\left( 2\right) }K_{\phantom{a}\mu }^{a%
\phantom{\mu}b}\right) \right\vert _{\mathrm{BPS}}\,, \\
\left. \left( \delta ^{\left( 2\right) }\Omega ^{abc}\right) \right\vert _{%
\mathrm{BPS}}=& \left[ \left. \left( \delta ^{\left( 2\right) }e^{\mu
a}\right) \right\vert _{\mathrm{BPS}}e^{\nu b}+e^{\mu a}\left. \left( \delta
^{\left( 2\right) }e^{\nu b}\right) \right\vert _{\mathrm{BPS}}\right]
\left( \partial _{\mu }e_{\nu }^{c}-\partial _{\nu }e_{\mu }^{c}\right) +
\notag \\
& +e^{\mu a}e^{\nu b}\left[ \partial _{\mu }\left. \left( \delta ^{\left(
2\right) }e_{\nu }^{c}\right) \right\vert _{\mathrm{BPS}}-\partial _{\nu
}\left. \left( \delta ^{\left( 2\right) }e_{\mu }^{c}\right) \right\vert _{%
\mathrm{BPS}}\right] \,, \\
\left. \left( \delta ^{\left( 2\right) }K_{\phantom{a}\mu }^{a\phantom{\mu}%
b}\right) \right\vert _{\mathrm{BPS}}=& \left. \left( \delta ^{\left(
1\right) }\bar{\psi}_{\rho }\right) \right\vert _{\mathrm{BPS}}e^{\rho \left[
a\right. }\gamma ^{\left. b\right] }\left. \left( \delta ^{\left( 1\right)
}\psi _{\mu }\right) \right\vert _{\mathrm{BPS}}+  \notag \\
& +\frac{1}{2}\left. \left( \delta ^{\left( 1\right) }\bar{\psi}_{\rho
}\right) \right\vert _{\mathrm{BPS}}\gamma _{\mu }\left. \left( \delta
^{\left( 1\right) }\psi _{\nu }\right) \right\vert _{\mathrm{BPS}}e^{\rho
a}e^{\nu b}\ .
\end{align}

\subsection{\label{54}Fourth Order}

Finally, at the fourth order, by using the identity \cite{Gunaydin:1983bi}
\end{subequations}
\begin{equation*}
h^{I}h_{Ix}=0\ ,
\end{equation*}%
one achieves the following results :
\begin{subequations}
\begin{align}
\left. \left( \delta ^{\left( 4\right) }e_{\mu }^{a}\right) \right\vert _{%
\mathrm{BPS}}=& \frac{1}{2}\bar{\epsilon}\gamma ^{a}\left. \left( \delta
^{\left( 3\right) }\psi _{\mu }\right) \right\vert _{\mathrm{BPS}}\neq 0\,,
\\
\left. \left( \delta ^{\left( 4\right) }\phi ^{x}\right) \right\vert _{%
\mathrm{BPS}}=& ~0\,, \\
\left. \left( \delta ^{\left( 4\right) }A_{\mu }^{I}\right) \right\vert _{%
\mathrm{BPS}}=& -\frac{i}{2}\sqrt{\frac{3}{2}}\bar{\epsilon}h^{I}\left.
\left( \delta ^{\left( 3\right) }\psi _{\mu }\right) \right\vert _{\mathrm{%
BPS}}\neq 0\ .
\end{align}%
Once again, the supercovariant field strength, the covariant derivative on $%
\phi ^{x}$ and the spin connection $\omega _{\mu }^{ab}$ all vanish.

\section{\label{Conclusion}Conclusion}

The general structure of the fermionic wigging (\ref{expansion0}) along a $4$%
-component anti-Killing spinor,~as well as the results reported in Secs.~\ref%
{52} and \ref{54}, do imply that the attractor values of the real scalar
fields $\phi ^{x}$ in the near--horizon $AdS_{2}\times S^{3}$ geometry of
the $1/2$--BPS extremal (electric) black hole are \textit{not} corrected by
the fermionic wigging itself; an analogous result holds for extremal
(magnetic) black string with a near horizon geometry $AdS_{3}\times S^{2}$ (%
\textit{cfr. e.g.} \cite{Kraus:2005gh} and \cite{Ceresole:2007rq}).

Thus, the attractor values of the scalar fields $\phi ^{x}$ are still fixed
purely in terms of the black hole (electric) charges :
\end{subequations}
\begin{align}
\left. \widehat{\phi }^{x}\right\vert _{\mathrm{BPS}}=& \left. \left(
e^{\delta }\phi \right) \right\vert _{\mathrm{BPS}}=  \notag \\
=& \left. \phi ^{x}\right\vert _{BPS}+\left. \left( \delta ^{\left( 1\right)
}\phi ^{x}\right) \right\vert _{\mathrm{BPS}}+\frac{1}{2!}\left. \left(
\delta ^{\left( 2\right) }\phi ^{x}\right) \right\vert _{\mathrm{BPS}}+
\notag \\
& +\frac{1}{3!}\left. \left( \delta ^{\left( 3\right) }\phi ^{x}\right)
\right\vert _{\mathrm{BPS}}+\frac{1}{4!}\left. \left( \delta ^{\left(
4\right) }\phi ^{x}\right) \right\vert _{\mathrm{BPS}}  \notag \\
=& \left. \phi \right\vert _{\mathrm{BPS}}\,,  \label{final0}
\end{align}%
as it holds for the attractor mechanism on the purely bosonic background (%
\textit{cfr. e.g.} \cite{Chamseddine:1996pi,Chamseddine:1999qs,Kraus:2005gh}%
). It should also be stressed that the result (\ref{final0}) does \textit{not%
} depend on the specific data of the real special geometry of the manifold
defined by the scalars of the vector multiplets.\medskip

We would like to stress once again that we adopted the approximation of
computing the fermionic wig by performing a perturbation of the unwigged,
purely bosonic BPS extremal black hole solution while keeping the radius of
the event horizon unchanged.

The complete analysis of the fully-backreacted wigged black hole solution,
including the study of its thermodynamical properties and the computation of
its Bekenstein-Hawking entropy is left for future work. This study can also
be generalized to the non-supersymmetric (non-BPS) case\footnote{%
Note that in this case the series (\ref{expansion0}) truncates at the $8^{%
\mathrm{th}}$ order}.\medskip

It should also be remarked that in $D=4,$ the attractor mechanism receives a
priori \textit{non-vanishing} corrections from bilinear terms in the
anti-Killing spinor $\epsilon $ \cite{Gentile:2013gea}.

Further investigation of such an important difference concerning wig
corrections to the attractor mechanism in $D=4$ and $D=5$ is currently in
progress, and results will be reported elsewhere. Here, we confine ourselves
to anticipate that the aforementioned \textit{non-vanishing} wig corrections
in $D=4$ can be related to the intrinsically \textit{dyonic} nature of the
four-dimensional \textit{\textquotedblleft large"} charge configurations,
namely to the fact that charge configurations giving rise to a non-vanishing
area of the horizon, and thus to a well-defined attractor mechanism for
scalar dynamics, contain \textit{both} electric and magnetic charges.\medskip

As further venues of research, we finally would like to mention that
fermionic wigging techniques could also be applied to other asymptotically
flat $D=5$ solutions, such as black rings \cite{Black-Ring,Kraus:2005gh} and
\textquotedblleft black Saturns" \cite{Black-Saturn}, as well to extended $%
\mathcal{N}>2$ supergravity theories in five dimensions.

\section*{Acknowledgments}

We would like to thank Anna Ceresole and Gianguido Dall'Agata for useful discussions on black holes in supergravity and on real special geometry.

The work of LGCG is partially supported by MIUR grant RBFR10QS5J
\textquotedblleft String Theory and Fundamental
Interactions\textquotedblright .

The work of PAG is partially supported by the MIUR-PRIN contract 2009-KHZKRX.

The work of A. Marrani is supported in part by the FWO - Vlaanderen, Project
No. G.0651.11, and in part by the Interuniversity Attraction Poles Programme
initiated by the Belgian Science Policy (P7/37).

The work of A. Mezzalira is partially supported by IISN - Belgium
(conventions 4.4511.06 and 4.4514.08), by the \textquotedblleft Communaut%
\'{e} Fran\c{c}aise de Belgique" through the ARC program and by the ERC
through the \textquotedblleft SyDuGraM" Advanced Grant.

\appendix

\section{\label{App-Notations-Ids}Notation and Identities}

We follow the notations in \cite{Gunaydin:1983bi}. We adopt the Lorentzian $%
D=5$ metric signature $\left( -,+,+,+,+\right) $ and we consider
symplectic--Majorana spinors satisfying
\begin{equation}
\bar{\lambda}^{i}=(\lambda _{i})^{\dagger }\gamma ^{0}=\lambda ^{iT}\mathcal{%
C}\ ,  \label{symMajcond}
\end{equation}%
where the charge conjugation matrix $\mathcal{C}$ fulfills the condition
\begin{equation}
\mathcal{C}^{T}=-\mathcal{C}=\mathcal{C}^{-1}\ ,\qquad \mathcal{C}^{2}=-1\ ,
\end{equation}%
and
\begin{align}
& \mathcal{C}\gamma _{\mu }\mathcal{C}^{-1}=(\gamma _{\mu
})^{T}\Longrightarrow \mathcal{C}(\gamma _{\mu })^{T}\mathcal{C}=-\gamma
_{\mu }\ ,  \notag \\
& \mathcal{C}(\gamma _{\mu \nu })^{T}\mathcal{C}=\gamma _{\mu \nu }\ ,
\end{align}%
from which one obtains
\begin{equation*}
\left( \mathcal{C}\gamma _{\mu }\right) ^{T}=-\mathcal{C}\gamma _{\mu }\
,\qquad \left( \mathcal{C}\gamma _{\mu \nu }\right) ^{T}=\mathcal{C}\gamma
_{\mu \nu }\ .
\end{equation*}%
Notice that $\mathcal{C}$ and $\mathcal{C}\gamma _{\mu }$ are antisymmetric
matrices, while $\mathcal{C}\gamma _{\mu \nu }$ is a symmetric one.
Spinorial indices $i=1,2$ are raised and lowered as follows
\begin{equation*}
V^{i}=\varepsilon ^{ij}V_{j}\ ,\qquad V_{i}=V^{j}\varepsilon _{ji}\ ,
\end{equation*}%
with
\begin{equation*}
\varepsilon _{12}=\varepsilon ^{12}=1\ .
\end{equation*}%
From these relations, one can derive the following identities :
\begin{align}
\bar{\lambda}^{i}\chi _{i}& =\bar{\lambda}^{i}\chi ^{j}\varepsilon _{ji}=-%
\bar{\chi}^{i}\lambda _{i}=\bar{\chi}_{i}\lambda ^{i}\ , \\
\bar{\lambda}^{i}\gamma _{\mu }\chi _{i}& =\bar{\lambda}^{i}\gamma _{\mu
}\chi ^{j}\varepsilon _{ji}=-\bar{\chi}^{j}\gamma _{\mu }\lambda _{j}=\bar{%
\chi}_{i}\gamma _{\mu }\lambda ^{i}\ , \\
\bar{\lambda}^{i}\gamma _{\mu \nu }\chi _{i}& =\bar{\lambda}^{i}\gamma _{\mu
\nu }\chi ^{j}\varepsilon _{ji}=\bar{\chi}^{j}\gamma _{\mu \nu }\lambda
_{j}=-\bar{\chi}_{i}\gamma _{\mu \nu }\lambda ^{i}\ ,
\end{align}%
yielding
\begin{align}
\bar{\lambda}^{i}\lambda _{i}& =0\ , \\
\bar{\lambda}^{i}\gamma _{\mu }\lambda _{i}& =0\ , \\
\bar{\lambda}^{i}\gamma _{\mu \nu }\lambda _{i}& \neq 0\ .
\end{align}

\section{\label{3rd-Order}Third Order}

At third order in $\mathcal{N}=2$, $D=5$ supersymmetry iterated
transformations, one finds\footnote{$\nabla _{t}\nabla _{y}h_{x}^{I}$ can be
elaborated by exploiting Eq.~(\ref{covDer2}). Furthermore, $\nabla
_{w}\nabla _{u}T^{xyz}=12\widetilde{E}^{xyz}{}_{wu}$, where the rank--$5$
completely symmetric tensor $\widetilde{E}^{xyz}{}_{wu}$ is the real special
geometry analogue \cite{Cerchiai:2010xv} of the so--called $E$\textit{%
--tensor} of special K\"{a}hler geometry \cite{deWit:1992wf}; by using the
last of (\ref{jazz1}), a similar result holds for $\nabla _{u}\nabla
_{t}\Gamma _{yz}^{x}$.} {\footnotesize
\begin{align}
\left( \delta ^{\left( 3\right) }e_{\mu }^{a}\right) =& \frac{1}{2}\bar{%
\epsilon}\gamma ^{a}\left( \delta ^{\left( 2\right) }\psi _{\mu }\right) \,,
\\
\left( \delta ^{\left( 3\right) }\psi^i_{\mu }\right) =& \left( \delta
^{\left( 2\right) }{\mathcal{D}}_{\mu }\right) \epsilon^i -\frac{1}{6}%
\epsilon _{j}\bar{\lambda}^{ix}\gamma _{\mu }\left( \delta ^{\left( 2\right)
}\lambda _{x}^{j}\right) +\frac{1}{12}\gamma _{\mu \nu }\epsilon _{j}\bar{%
\lambda}^{ix}\gamma ^{\nu }\left( \delta ^{\left( 2\right) }\lambda
_{x}^{j}\right) +  \notag \\
& -\frac{1}{48}\gamma _{\mu \nu \rho }\epsilon _{j}\bar{\lambda}^{ix}\gamma
^{\nu \rho }\left( \delta ^{\left( 2\right) }\lambda _{x}^{j}\right) +\frac{1%
}{12}\gamma ^{\nu }\epsilon_j \bar{\lambda}^{ix}\gamma _{\mu \nu }\left(
\delta ^{\left( 2\right) }\lambda _{x}^{j}\right) +  \notag \\
& -\frac{1}{3}\left( \delta ^{\left( 1\right) }e_{\mu }^{a}\right) \epsilon
_{j}\bar{\lambda}^{ix}\gamma _{a}\left( \delta ^{\left( 1\right) }\lambda
_{x}^{j}\right) -\frac{1}{3}\epsilon _{j}\left( \delta ^{\left( 1\right) }%
\bar{\lambda}^{ix}\right) \gamma _{\mu }\left( \delta ^{\left( 1\right)
}\lambda _{x}^{j}\right) +  \notag \\
& +\frac{1}{6}\left( \delta ^{\left( 1\right) }e_{\mu }^{a}\right) \gamma
_{ab} \epsilon_j \bar{\lambda}^{ix}\gamma ^{b}\left( \delta ^{\left(
1\right) }\lambda _{x}^{j}\right) +\frac{1}{6}\gamma _{\mu \nu } \epsilon_j
\left( \delta ^{\left( 1\right) }\bar{\lambda}^{ix}\right) \gamma ^{\nu
}\left( \delta ^{\left( 1\right) }\lambda _{x}^{j}\right) +  \notag \\
& -\frac{1}{24}\left( \delta ^{\left( 1\right) }e_{\mu }^{a}\right) \gamma
_{abc}\epsilon _{j}\bar{\lambda}^{ix}\gamma ^{bc}\left( \delta ^{\left(
1\right) }\lambda _{x}^{j}\right) -\frac{1}{24}\gamma _{\mu \nu \rho
}\epsilon _{j}\left( \delta ^{\left( 1\right) }\bar{\lambda}^{ix}\right)
\gamma ^{\nu \rho }\left( \delta ^{\left( 1\right) }\lambda _{x}^{j}\right) +
\notag \\
& +\frac{1}{6}\left( \delta ^{\left( 1\right) }e_{\mu }^{a}\right) \gamma
_{ab}\epsilon _{j}\bar{\lambda}^{ix}\gamma ^{b}\left( \delta ^{\left(
1\right) }\lambda _{x}^{j}\right) +\frac{1}{6}\gamma _{\mu \nu }\epsilon
_{j}\left( \delta ^{\left( 1\right) }\bar{\lambda}^{ix}\right) \gamma ^{\nu
}\left( \delta ^{\left( 1\right) }\lambda _{x}^{j}\right) +  \notag \\
& -\frac{1}{6}\epsilon _{j}\bar{\lambda}^{ix}\gamma _{a}\lambda
_{x}^{j}\left( \delta ^{\left( 2\right) }e_{\mu }^{a}\right) -\frac{1}{3}%
\left( \delta ^{\left( 1\right) }e_{\mu }^{a}\right) \epsilon _{j}\left(
\delta ^{\left( 1\right) }\bar{\lambda}^{ix}\right) \gamma _{a}\lambda
_{x}^{j}+  \notag \\
& -\frac{1}{3}\epsilon _{j}\left( \delta ^{\left( 2\right) }\bar{\lambda}%
^{ix}\right) \gamma _{\mu }\lambda _{x}^{j}+\frac{1}{12}\gamma _{ab}\epsilon
_{j}\bar{\lambda}^{ix}\gamma ^{b}\lambda _{x}^{j}\left( \delta ^{\left(
2\right) }e_{\mu }^{a}\right)  \notag \\
& +\frac{1}{6}\gamma _{ab}\epsilon _{j}\left( \delta ^{\left( 1\right) }\bar{%
\lambda}^{ix}\right) \gamma ^{b}\lambda _{x}^{j}\left( \delta ^{\left(
1\right) }e_{\mu }^{a}\right) +\frac{1}{12}\gamma _{\mu \nu }\epsilon
_{j}\left( \delta ^{\left( 2\right) }\bar{\lambda}^{ix}\right) \gamma ^{\nu
}\lambda _{x}^{j}+  \notag \\
& -\frac{1}{48}\gamma _{abc}\epsilon _{j}\bar{\lambda}^{ix}\gamma
^{bc}\lambda _{x}^{j}\left( \delta ^{\left( 2\right) }e_{\mu }^{a}\right) -%
\frac{1}{24}\gamma _{abc}\epsilon _{j}\left( \delta ^{\left( 1\right) }\bar{%
\lambda}^{ix}\right) \gamma ^{bc}\lambda _{x}^{j}\left( \delta ^{\left(
1\right) }e_{\mu }^{a}\right) +  \notag \\
& -\frac{1}{48}\gamma _{\mu \nu \rho }\epsilon _{j}\left( \delta ^{\left(
2\right) }\bar{\lambda}^{ix}\right) \gamma ^{\nu \rho }\lambda _{x}^{j}+%
\frac{1}{12}\gamma ^{\nu }\epsilon _{j}\left( \delta ^{\left( 2\right) }\bar{%
\lambda}^{ix}\right) \gamma _{\mu \nu }\lambda _{x}^{j}  \notag \\
& +\frac{1}{12}\gamma ^{b}\epsilon _{j}\bar{\lambda}^{ix}\gamma _{ab}\lambda
_{x}^{j}\left( \delta ^{\left( 2\right) }e_{\mu }^{a}\right) +\frac{1}{6}%
\gamma ^{b}\epsilon _{j}\left( \delta ^{\left( 1\right) }\bar{\lambda}%
^{ix}\right) \gamma _{ab}\lambda _{x}^{j}\left( \delta ^{\left( 1\right)
}e_{\mu }^{a}\right) +  \notag \\
& +\frac{i}{4\sqrt{6}}h_{I}\tilde{F}_{\nu \rho }^{I}\left[ \left( \delta
^{\left( 2\right) }e_{\mu }^{a}\right) \epsilon _{b}^{\nu }e_{c}^{\rho
}+e_{\mu }^{a}\left( \delta ^{\left( 2\right) }\epsilon _{b}^{\nu }\right)
e_{c}^{\rho }+e_{\mu }^{a}\epsilon _{b}^{\nu }\left( \delta ^{\left(
2\right) }e_{c}^{\rho }\right) +\right.  \notag \\
& \left. +2\left( \delta ^{\left( 1\right) }e_{\mu }^{a}\right) \left(
\delta ^{\left( 1\right) }\epsilon _{b}^{\nu }\right) e_{c}^{\rho }+2\left(
\delta ^{\left( 1\right) }e_{\mu }^{a}\right) \epsilon _{b}^{\nu }\left(
\delta ^{\left( 1\right) }e_{c}^{\rho }\right) +2e_{\mu }^{a}\left( \delta
^{\left( 1\right) }\epsilon _{b}^{\nu }\right) \left( \delta ^{\left(
1\right) }e_{c}^{\rho }\right) \right] \left( \gamma _{a}^{\phantom{a}%
bc}-4\delta _{a}^{b}\gamma ^{c}\right) \epsilon ^{i}+  \notag \\
& +\frac{i}{6}h_{Iz}\left( \delta ^{\left( 1\right) }\phi ^{z}\right) \tilde{%
F}_{\nu \rho }^{I}\left[ \left( \delta ^{\left( 1\right) }e_{\mu
}^{a}\right) e_{b}^{\nu }e_{c}^{\rho }+e_{\mu }^{a}\left( \delta ^{\left(
1\right) }e_{b}^{\nu }\right) e_{c}^{\rho }+e_{\mu }^{a}e_{b}^{\nu }\left(
\delta ^{\left( 1\right) }e_{c}^{\rho }\right) \right] \left( \gamma _{a}^{%
\phantom{a}bc}-4\delta _{a}^{b}\gamma ^{c}\right) \epsilon ^{i}+  \notag \\
& +\frac{i}{12}h_{Iz}\left( \delta ^{\left( 2\right) }\phi ^{z}\right)
\tilde{F}_{\nu \rho }^{I}\left( \gamma _{\mu }^{\phantom{\m}\nu \rho
}-4\delta _{\mu }^{\nu }\gamma ^{\rho }\right) \epsilon ^{i}+  \notag \\
& +\frac{i}{12}\nabla _{y}h_{Iz}\left( \delta ^{\left( 1\right) }\phi
^{z}\right) \left( \delta ^{\left( 1\right) }\phi ^{y}\right) \tilde{F}_{\nu
\rho }^{I}\left( \gamma _{\mu }^{\phantom{\m}\nu \rho }-4\delta _{\mu }^{\nu
}\gamma ^{\rho }\right) \epsilon ^{i}+  \notag \\
& +\frac{i}{2\sqrt{6}}h_{I}\left( \delta ^{\left( 1\right) }\tilde{F}_{\nu
\rho }^{I}\right) \left[ \left( \delta ^{\left( 1\right) }e_{\mu
}^{a}\right) e_{b}^{\nu }e_{c}^{\rho }+e_{\mu }^{a}\left( \delta ^{\left(
1\right) }e_{b}^{\nu }\right) e_{c}^{\rho }+e_{\mu }^{a}e_{b}^{\nu }\left(
\delta ^{\left( 1\right) }e_{c}^{\rho }\right) \right] \left( \gamma _{a}^{%
\phantom{a}bc}-4\delta _{a}^{b}\gamma ^{c}\right) \epsilon ^{i}+  \notag \\
& +\frac{i}{6}h_{Iz}\left( \delta ^{\left( 1\right) }\phi ^{z}\right) \left(
\delta ^{\left( 1\right) }\tilde{F}_{\nu \rho }^{I}\right) \left( \gamma
_{\mu }^{\phantom{\m}\nu \rho }-4\delta _{\mu }^{\nu }\gamma ^{\rho }\right)
\epsilon ^{i}  \notag \\
& +\frac{i}{4\sqrt{6}}h_{I}\left( \delta ^{\left( 2\right) }\tilde{F}_{\nu
\rho }^{I}\right) \left( \gamma _{\mu }^{\phantom{\m}\nu \rho }-4\delta
_{\mu }^{\nu }\gamma ^{\rho }\right) \epsilon ^{i} \,, \\
\left( \delta ^{\left( 3\right) }\phi ^{x}\right) =& \frac{i}{2}\bar{\epsilon%
}\left( \delta ^{\left( 2\right) }\lambda ^{x}\right) \,, \\
\left( \delta ^{\left( 3\right) }A_{\mu }^{I}\right) =& -\frac{1}{2}\bar{%
\epsilon}\gamma _{\mu }\left( \delta ^{\left( 2\right) }\lambda ^{x}\right)
h_{x}^{I}-\frac{1}{2}\left( \delta ^{\left( 2\right) }e_{\mu }^{a}\right)
\bar{\epsilon}\gamma _{a}\lambda ^{x}h_{x}^{I}+  \notag \\
& -\frac{i}{2}\sqrt{\frac{3}{2}}\bar{\epsilon}h^{I}\left( \delta ^{\left(
2\right) }\psi _{\mu }\right) +\frac{i}{2}h_{x}^{I}\left( \delta ^{\left(
2\right) }\phi ^{x}\right) \bar{\epsilon}\psi _{\mu }+  \notag \\
& -\frac{1}{2}\bar{\epsilon}\gamma _{\mu }\lambda ^{x}\nabla
_{y}h_{x}^{I}\left( \delta ^{\left( 2\right) }\phi ^{y}\right) +  \notag \\
& +ih_{x}^{I}\left( \delta ^{\left( 1\right) }\phi ^{x}\right) \bar{\epsilon}%
\left( \delta ^{\left( 1\right) }\psi _{\mu }\right) +\frac{i}{2}\nabla
_{y}h_{x}^{I}\left( \delta ^{\left( 1\right) }\phi ^{y}\right) \left( \delta
^{\left( 1\right) }\phi ^{x}\right) \bar{\epsilon}\psi _{\mu }+  \notag \\
& -\left( \delta ^{\left( 1\right) }e_{\mu }^{a}\right) \bar{\epsilon}\gamma
_{a}\left( \delta ^{\left( 1\right) }\lambda ^{x}\right) h_{x}^{I}-\left(
\delta ^{\left( 1\right) }e_{\mu }^{a}\right) \bar{\epsilon}\gamma
_{a}\lambda ^{x}\nabla _{y}h_{x}^{I}\left( \delta ^{\left( 1\right) }\phi
^{y}\right) +  \notag \\
& -\bar{\epsilon}\gamma _{\mu }\left( \delta ^{\left( 1\right) }\lambda
^{x}\right) \nabla _{y}h_{x}^{I}\left( \delta ^{\left( 1\right) }\phi
^{y}\right) -\frac{1}{2}\bar{\epsilon}\gamma _{\mu }\lambda ^{x}\nabla
_{t}\nabla _{y}h_{x}^{I}\left( \delta ^{\left( 1\right) }\phi ^{y}\right)
\left( \delta ^{\left( 1\right) }\phi ^{t}\right) \,, \\
\left( \delta ^{\left( 3\right) }\lambda ^{ix}\right) =& -\left( \delta
^{\left( 1\right) }e_{a}^{\mu }\right) \gamma ^{a}\left( \delta ^{\left(
1\right) }\widehat{\mathcal{D}}_{\mu }\phi ^{x}\right) \epsilon ^{i}-\frac{i%
}{2}\left( \delta ^{\left( 2\right) }e_{a}^{\mu }\right) \gamma ^{a}\widehat{%
\mathcal{D}}_{\mu }\phi ^{x}\epsilon ^{i}+  \notag \\
& -\frac{i}{2}\gamma ^{\mu }\left( \delta ^{\left( 2\right) }\widehat{%
\mathcal{D}}_{\mu }\phi ^{x}\right) \epsilon ^{i}-\frac{1}{4\sqrt{6}}%
T^{xyz}\gamma ^{\mu }\epsilon _{j}\bar{\lambda}_{y}^{i}\gamma _{\mu }\left(
\delta ^{\left( 2\right) }\lambda _{z}^{j}\right) +  \notag \\
& +\frac{1}{4}\sqrt{\frac{3}{2}}T^{xyz}\epsilon _{j}\bar{\lambda}%
_{y}^{i}\left( \delta ^{\left( 2\right) }\lambda _{z}^{j}\right) -\frac{1}{8%
\sqrt{6}}T^{xyz}\gamma ^{\mu \nu }\epsilon _{j}\bar{\lambda}_{y}^{i}\gamma
_{\mu \nu }\left( \delta ^{\left( 2\right) }\lambda _{z}^{j}\right) +  \notag
\\
& -2\left( \delta ^{\left( 2\right) }\phi ^{y}\right) \Gamma _{yz}^{x}\left(
\delta ^{\left( 1\right) }\lambda ^{zi}\right) +\frac{1}{2}\sqrt{\frac{3}{2}}%
T^{xyz}\epsilon _{j}\left( \delta ^{\left( 1\right) }\bar{\lambda}%
_{y}^{i}\right) \left( \delta ^{\left( 1\right) }\lambda _{z}^{j}\right) +
\notag \\
& -\frac{1}{4\sqrt{6}}T^{xyz}\gamma ^{\mu \nu }\epsilon _{j}\left( \delta
^{\left( 1\right) }\bar{\lambda}_{y}^{i}\right) \gamma _{\mu \nu }\left(
\delta ^{\left( 1\right) }\lambda _{z}^{j}\right) -\frac{1}{2\sqrt{6}}%
T^{xyz}\gamma ^{\mu }\epsilon _{j}\left( \delta ^{\left( 1\right) }\bar{%
\lambda}_{y}^{i}\right) \gamma _{\mu }\left( \delta ^{\left( 1\right)
}\lambda _{z}^{j}\right) +  \notag \\
& -\left( \delta ^{\left( 1\right) }\phi ^{y}\right) \Gamma _{yz}^{x}\left(
\delta ^{\left( 2\right) }\lambda ^{zi}\right) -\left( \delta ^{\left(
3\right) }\phi ^{y}\right) \Gamma _{yz}^{x}\lambda ^{zi}+  \notag \\
& +\frac{1}{4}\sqrt{\frac{3}{2}}T^{xyz}\epsilon _{j}\left( \delta ^{\left(
2\right) }\bar{\lambda}_{y}^{i}\right) \lambda _{z}^{j}-\frac{1}{8\sqrt{6}}%
T^{xyz}\gamma ^{\mu \nu }\epsilon _{j}\left( \delta ^{\left( 2\right) }\bar{%
\lambda}_{y}^{i}\right) \gamma _{\mu \nu }\lambda _{z}^{j}+  \notag \\
& +\frac{1}{2}\sqrt{\frac{3}{2}}\nabla _{u}T^{xyz}\left( \delta ^{\left(
1\right) }\phi ^{u}\right) \epsilon _{j} \bar{\lambda}_{y}^{i}\left( \delta
^{\left( 1\right) }\lambda _{z}^{i}\right) -\frac{1}{4\sqrt{6}}\nabla
_{u}T^{xyz}\left( \delta ^{\left( 1\right) }\phi ^{u}\right) \gamma ^{\mu
\nu }\epsilon _{j}\left( \delta ^{\left( 1\right) }\bar{\lambda}%
_{y}^{i}\right) \gamma _{\mu \nu }\lambda _{z}^{j}+  \notag \\
& -\frac{1}{2\sqrt{6}}\nabla _{u}T^{xyz}\left( \delta ^{\left( 1\right)
}\phi ^{u}\right) \gamma ^{\mu }\epsilon _{j}\bar{\lambda}_{y}^{i}\gamma
_{\mu }\left( \delta ^{\left( 1\right) }\lambda _{z}^{j}\right) +\frac{1}{2}%
\sqrt{\frac{3}{2}}\nabla _{u}T^{xyz}\left( \delta ^{\left( 1\right) }\phi
^{u}\right) \epsilon _{j}\left( \delta ^{\left( 1\right) }\bar{\lambda}%
_{y}^{i}\right) \lambda _{z}^{j}+  \notag \\
& -\frac{1}{2\sqrt{6}}\nabla _{u}T^{xyz}\left( \delta ^{\left( 1\right)
}\phi ^{u}\right) \gamma ^{\mu }\epsilon _{j}\left( \delta ^{\left( 1\right)
}\bar{\lambda}_{y}^{i}\right) \gamma _{\mu }\lambda _{z}^{j}-2\left( \delta
^{\left( 1\right) }\phi ^{y}\right) \nabla _{t}\Gamma _{yz}^{x}\left( \delta
^{\left( 1\right) }\phi ^{t}\right) \left( \delta ^{\left( 1\right) }\lambda
^{zi}\right) +  \notag \\
& -2\left( \delta ^{\left( 2\right) }\phi ^{y}\right) \nabla _{t}\Gamma
_{yz}^{x}\left( \delta ^{\left( 1\right) }\phi ^{t}\right) \lambda ^{zi}+%
\frac{1}{4}\sqrt{\frac{3}{2}}\nabla _{u}T^{xyz}\left( \delta ^{\left(
2\right) }\phi ^{u}\right) \epsilon _{j}\bar{\lambda}_{y}^{i}\lambda
_{z}^{j}+  \notag \\
& +\frac{1}{4}\sqrt{\frac{3}{2}}\nabla _{w}\nabla _{u}T^{xyz}\left( \delta
^{\left( 1\right) }\phi ^{u}\right) \left( \delta ^{\left( 1\right) }\phi
^{w}\right) \epsilon _{j}\bar{\lambda}_{y}^{i}\lambda _{z}^{j}+  \notag \\
& -\frac{1}{8\sqrt{6}}\nabla _{w}\nabla _{u}T^{xyz}\left( \delta ^{\left(
1\right) }\phi ^{u}\right) \left( \delta ^{\left( 1\right) }\phi ^{w}\right)
\gamma ^{\mu \nu }\epsilon _{j}\bar{\lambda}_{y}^{i}\gamma _{\mu \nu
}\lambda _{z}^{j}+  \notag \\
& -\frac{1}{4\sqrt{6}}\nabla _{w}\nabla _{u}T^{xyz}\left( \delta ^{\left(
1\right) }\phi ^{u}\right) \left( \delta ^{\left( 1\right) }\phi ^{w}\right)
\gamma ^{\mu }\epsilon _{j}\bar{\lambda}_{y}^{i}\gamma _{\mu }\lambda
_{z}^{j}+  \notag \\
& -2\left( \delta ^{\left( 1\right) }\phi ^{y}\right) \nabla _{u}\nabla
_{t}\Gamma _{yz}^{x}\left( \delta ^{\left( 1\right) }\phi ^{t}\right) \left(
\delta ^{\left( 1\right) }\phi ^{u}\right) \lambda ^{zi}+  \notag \\
& +\frac{1}{4}\gamma \cdot \tilde{F}^{I}\nabla _{t}h_{I}^{x}\left( \delta
^{\left( 2\right) }\phi ^{t}\right) \epsilon ^{i}+\frac{1}{2}\gamma \cdot
\left( \delta ^{\left( 1\right) }\tilde{F}^{I}\right) \nabla
_{t}h_{I}^{x}\left( \delta ^{\left( 1\right) }\phi ^{t}\right) \epsilon ^{i}+
\notag \\
& +\frac{1}{4}\gamma \cdot \left( \delta ^{\left( 2\right) }\tilde{F}%
^{I}\right) h_{I}^{x}\epsilon ^{i}+\frac{1}{4}\gamma \cdot \tilde{F}%
^{I}\nabla _{u}\nabla _{t}h_{I}^{x}\left( \delta ^{\left( 1\right) }\phi
^{t}\right) \left( \delta ^{\left( 1\right) }\phi ^{u}\right) \epsilon ^{i}+
\notag \\
& +\frac{1}{4}\gamma ^{ab}\left[ \left( \delta ^{\left( 2\right) }e_{a}^{\mu
}\right) e_{b}^{\nu }+2\left( \delta ^{\left( 1\right) }e_{a}^{\mu }\right)
\left( \delta ^{\left( 1\right) }e_{b}^{\nu }\right) +e_{a}^{\mu }\left(
\delta ^{\left( 2\right) }e_{b}^{\nu }\right) \right] \tilde{F}_{\mu \nu
}h_{I}^{x}+  \notag \\
& +\gamma ^{ab} \left( \delta ^{\left( 1\right) }e_{a}^{\mu }\right)
e_{b}^{\nu } \left( \delta ^{\left( 1\right) }\tilde{F}_{\mu \nu }\right)
h_{I}^{x}+  \notag \\
& +\gamma ^{ab} \left( \delta ^{\left( 1\right) }e_{a}^{\mu }\right)
e_{b}^{\nu } \tilde{F}_{\mu \nu }\nabla _{t}h_{I}^{x}\left( \delta ^{\left(
1\right) }\phi ^{t}\right) +  \notag \\
& -\left( \delta ^{\left( 1\right) }\phi ^{y}\right) \nabla _{t}\Gamma
_{yz}^{x}\left( \delta ^{\left( 2\right) }\phi ^{t}\right) \lambda ^{zi}-%
\frac{1}{4\sqrt{6}}\nabla _{u}T^{xyz}\left( \delta ^{\left( 2\right) }\phi
^{u}\right) \gamma ^{\mu }\epsilon _{j}\bar{\lambda}_{y}^{i}\gamma _{\mu
}\lambda _{z}^{j}+  \notag \\
& -\frac{1}{8\sqrt{6}}\nabla _{u}T^{xyz}\left( \delta ^{\left( 2\right)
}\phi ^{u}\right) \gamma ^{\mu \nu }\epsilon _{j}\bar{\lambda}_{y}^{i}\gamma
_{\mu \nu }\lambda _{z}^{j} \,,
\end{align}%
}with {\footnotesize
\begin{align}
\left( \delta ^{\left( 2\right) }\tilde{F}_{\mu \nu }^{I}\right) =& \left(
\delta ^{\left( 2\right) }\mathcal{F}_{\mu \nu }^{I}\right) +2\left( \delta
^{\left( 1\right) }\bar{\psi}_{\left[ \mu \right. }\right) \gamma _{\left.
\nu \right] }\left( \delta ^{\left( 1\right) }\lambda ^{x}\right)
h_{x}^{I}+2\left( \delta ^{\left( 1\right) }\bar{\psi}_{\left[ \mu \right.
}\right) \gamma _{\left. \nu \right] }\lambda ^{x}\nabla _{y}h_{x}^{I}\left(
\delta ^{\left( 1\right) }\phi ^{y}\right) +  \notag \\
& +2\bar{\psi}_{\left[ \mu \right. }\gamma _{\left. \nu \right] }\left(
\delta ^{\left( 1\right) }\lambda ^{x}\right) \nabla _{y}h_{x}^{I}\left(
\delta ^{\left( 1\right) }\phi ^{y}\right) +\left( \delta ^{\left( 2\right) }%
\bar{\psi}_{\left[ \mu \right. }\right) \gamma _{\left. \nu \right] }\lambda
^{x}h_{x}^{I}+  \notag \\
& +\bar{\psi}_{\left[ \mu \right. }\gamma _{\left. \nu \right] }\left(
\delta ^{\left( 2\right) }\lambda ^{x}\right) h_{x}^{I}+\bar{\psi}_{\left[
\mu \right. }\gamma _{\left. \nu \right] }\lambda ^{x}\nabla _{z}\nabla
_{y}h_{x}^{I}\left( \delta ^{\left( 1\right) }\phi ^{y}\right) \left( \delta
^{\left( 1\right) }\phi ^{z}\right) +  \notag \\
& +\bar{\psi}_{\left[ \mu \right. }\gamma _{\left. \nu \right] }\lambda
^{x}\nabla _{y}h_{x}^{I}\left( \delta ^{\left( 2\right) }\phi ^{y}\right) +%
\frac{i}{2}\sqrt{\frac{3}{2}}\left( \delta ^{\left( 2\right) }\bar{\psi}%
_{\mu }\right) \psi _{\nu }h^{I}+  \notag \\
& +\frac{i}{2}\sqrt{\frac{3}{2}}\bar{\psi}_{\mu }\left( \delta ^{\left(
2\right) }\psi _{\nu }\right) h^{I}+i\sqrt{\frac{3}{2}}\left( \delta
^{\left( 1\right) }\bar{\psi}_{\mu }\right) \left( \delta ^{\left( 1\right)
}\psi _{\nu }\right) h^{I}+  \notag \\
& -i\bar{\psi}_{\mu }\left( \delta ^{\left( 1\right) }\psi _{\nu }\right)
h_{x}^{I}\left( \delta ^{\left( 1\right) }\phi ^{x}\right) -\frac{i}{2}\bar{%
\psi}_{\mu }\psi _{\nu }h_{x}^{I}\left( \delta ^{\left( 2\right) }\phi
^{x}\right) +  \notag \\
& -i\left( \delta ^{\left( 1\right) }\bar{\psi}_{\mu }\right) \psi _{\nu
}h_{x}^{I}\left( \delta ^{\left( 1\right) }\phi ^{x}\right) -\frac{i}{2}\bar{%
\psi}_{\mu }\psi _{\nu }\nabla _{y}h_{x}^{I}\left( \delta ^{\left( 1\right)
}\phi ^{x}\right) \left( \delta ^{\left( 1\right) }\phi ^{y}\right) +  \notag
\\
& +2\left( \delta ^{\left( 1\right) }\bar{\psi}_{\left[ \mu \right. }\right)
\left( \delta ^{\left( 1\right) }e_{\left. \nu \right] }^{a}\right) \gamma
_{a}\lambda ^{x}h_{x}^{I}+\bar{\psi}_{\left[ \mu \right. }\left( \delta
^{\left( 2\right) }e_{\left. \nu \right] }^{a}\right) \gamma _{a}\lambda
^{x}h_{x}^{I}+  \notag \\
& +2\bar{\psi}_{\left[ \mu \right. }\left( \delta ^{\left( 1\right)
}e_{\left. \nu \right] }^{a}\right) \gamma _{a}\left( \delta ^{\left(
1\right) }\lambda ^{x}\right) h_{x}^{I}+2\bar{\psi}_{\left[ \mu \right.
}\left( \delta ^{\left( 1\right) }e_{\left. \nu \right] }^{a}\right) \gamma
_{a}\lambda ^{x}\nabla _{t}h_{x}^{I}\left( \delta ^{\left( 1\right) }\phi
^{t}\right) \,, \\
\left( \delta ^{\left( 2\right) }{\mathcal{D}}_{\mu }\right) =& \frac{1}{4}%
\left( \delta ^{\left( 2\right) }\omega _{\mu }^{ab}\right) \gamma _{ab} \,,
\\
\left( \delta ^{\left( 2\right) }\omega _{\mu }^{ab}\right) =& \frac{1}{2}%
\left( \delta ^{\left( 2\right) }e_{c\mu }\right) \left( \Omega
^{abc}-\Omega ^{bca}-\Omega ^{cab}\right) +\left( \delta ^{\left( 1\right)
}e_{c\mu }\right) \left[ \left( \delta ^{\left( 1\right) }\Omega
^{abc}\right) -\left( \delta ^{\left( 1\right) }\Omega ^{bca}\right) -\left(
\delta ^{\left( 1\right) }\Omega ^{cab}\right) \right] +  \notag \\
& +\frac{1}{2} e_{c\mu} \left[ \left( \delta ^{\left( 2\right) }\Omega
^{abc}\right) -\left( \delta ^{\left( 2\right) }\Omega ^{bca}\right) -\left(
\delta ^{\left( 2\right) }\Omega ^{cab}\right) \right] +\left( \delta
^{\left( 2\right) }K_{\phantom{a}\mu }^{a\phantom{\mu}b}\right) \,, \\
\left( \delta ^{\left( 2\right) }\Omega ^{abc}\right) =& \left[ \left(
\delta ^{\left( 2\right) }e^{\mu a}\right) e^{\nu b}+2\left( \delta ^{\left(
1\right) }e^{\mu a}\right) \left( \delta ^{\left( 1\right) }e^{\nu b}\right)
+e^{\mu a}\left( \delta ^{\left( 2\right) }e^{\nu b}\right) \right] \left(
\partial _{\mu }e_{\nu }^{c}-\partial _{\nu }e_{\mu }^{c}\right) +  \notag \\
& +2\left[ \left( \delta ^{\left( 1\right) }e^{\mu a}\right) e^{\nu
b}+e^{\mu a}\left( \delta ^{\left( 1\right) }e^{\nu b}\right) \right] \left[
\partial _{\mu }\left( \delta ^{\left( 1\right) }e_{\nu }^{c}\right)
-\partial _{\nu }\left( \delta ^{\left( 1\right) }e_{\mu }^{c}\right) \right]
+  \notag \\
& +e^{\mu a}e^{\nu b}\left[ \partial _{\mu }\left( \delta ^{\left( 2\right)
}e_{\nu }^{c}\right) -\partial _{\nu }\left( \delta ^{\left( 2\right)
}e_{\mu }^{c}\right) \right] \,, \\
\left( \delta ^{\left( 2\right) }K_{\phantom{a}\mu }^{a\phantom{\mu}%
b}\right) =& \frac{1}{2}\left[ \left( \delta ^{\left( 2\right) }\bar{\psi}%
_{\rho }\right) e^{\rho \left[ a\right. }\gamma ^{\left. b\right] }\psi
_{\mu }+2\left( \delta ^{\left( 1\right) }\bar{\psi}_{\rho }\right) \left(
\delta ^{\left( 1\right) }e^{\rho \left[ a\right. }\right) \gamma ^{\left. b%
\right] }\psi _{\mu }+\right.  \notag \\
& +2\left( \delta ^{\left( 1\right) }\bar{\psi}_{\rho }\right) e^{\rho \left[
a\right. }\gamma ^{\left. b\right] }\left( \delta ^{\left( 1\right) }\psi
_{\mu }\right) +\bar{\psi}_{\rho }\left( \delta ^{\left( 2\right) }e^{\rho %
\left[ a\right. }\right) \gamma ^{\left. b\right] }\psi _{\mu }+  \notag \\
& +2\bar{\psi}_{\rho }\left( \delta ^{\left( 1\right) }e^{\rho \left[
a\right. }\right) \gamma ^{\left. b\right] }\left( \delta ^{\left( 1\right)
}\psi _{\mu }\right) +\bar{\psi}^{\left[ a\right. }\gamma ^{\left. b\right]
}\left( \delta ^{\left( 2\right) }\psi _{\mu }\right) +  \notag \\
& +\frac{1}{2}\left( \delta ^{\left( 2\right) }\bar{\psi}_{\rho }\right)
e^{\rho a}\gamma _{\mu }\psi ^{b}+\left( \delta ^{\left( 1\right) }\bar{\psi}%
_{\rho }\right) \gamma _{\mu }\left( \delta ^{\left( 1\right) }\psi _{\nu
}\right) e^{\rho a}e^{\nu b}+  \notag \\
& +\left( \delta ^{\left( 1\right) }\bar{\psi}_{\rho }\right) \gamma _{\mu
}\psi ^{b}\left( \delta ^{\left( 1\right) }e^{\rho a}\right) +\left( \delta
^{\left( 1\right) }\bar{\psi}_{\rho }\right) \gamma _{c}\psi ^{b}e^{\rho
a}\left( \delta ^{\left( 1\right) }e_{\mu }^{c}\right) +  \notag \\
& +\left( \delta ^{\left( 1\right) }\bar{\psi}_{\rho }\right) \gamma _{\mu
}\psi _{\nu }e^{\rho a}\left( \delta ^{\left( 1\right) }e^{\nu b}\right) +%
\frac{1}{2}\bar{\psi}^{a}\gamma _{\mu }\left( \delta ^{\left( 2\right) }\psi
_{\nu }\right) e^{\nu b}+  \notag \\
& +\bar{\psi}_{\rho }\gamma _{\mu }\left( \delta ^{\left( 1\right) }\psi
_{\nu }\right) \left( \delta ^{\left( 1\right) }e^{\rho a}\right) e^{\nu b}+%
\bar{\psi}^{a}\gamma _{c}\left( \delta ^{\left( 1\right) }\psi _{\nu
}\right) \left( \delta ^{\left( 1\right) }e_{\mu }^{c}\right) e^{\nu b}+
\notag \\
& +\bar{\psi}^{a}\gamma _{\mu }\left( \delta ^{\left( 1\right) }\psi _{\nu
}\right) \left( \delta ^{\left( 1\right) }e^{\nu b}\right) +\frac{1}{2}\bar{%
\psi}_{\rho }\gamma _{\mu }\psi ^{b}\left( \delta ^{\left( 2\right) }e^{\rho
a}\right) +  \notag \\
& +\bar{\psi}_{\rho }\gamma _{c}\psi ^{b}\left( \delta ^{\left( 1\right)
}e^{\rho a}\right) \left( \delta ^{\left( 1\right) }e_{\mu }^{c}\right) +%
\bar{\psi}_{\rho }\gamma _{\mu }\psi _{\nu }\left( \delta ^{\left( 1\right)
}e^{\rho a}\right) \left( \delta ^{\left( 1\right) }e^{\nu b}\right) +
\notag \\
& +\frac{1}{2}\bar{\psi}^{a}\gamma _{c}\psi ^{b}\left( \delta ^{\left(
2\right) }e_{\mu }^{c}\right) +\bar{\psi}^{a}\gamma _{c}\psi _{\nu }\left(
\delta ^{\left( 1\right) }e_{\mu }^{c}\right) \left( \delta ^{\left(
1\right) }e^{\nu b}\right) +  \notag \\
& \left. +\frac{1}{2}\bar{\psi}^{a}\gamma _{\mu }\psi _{\nu }\left( \delta
^{\left( 2\right) }e^{\nu b}\right) \right] \,, \\
\left( \delta ^{\left( 2\right) }\widehat{{\mathcal{D}}}_{\mu }\phi
^{x}\right) =& \partial _{\mu }\left( \delta ^{\left( 2\right) }\phi
^{x}\right) -\frac{i}{2}\left( \delta ^{\left( 2\right) }\bar{\psi}_{\mu
}\right) \lambda ^{x}-i\left( \delta ^{\left( 1\right) }\bar{\psi}_{\mu
}\right) \left( \delta ^{\left( 1\right) }\lambda ^{x}\right) +  \notag \\
& -\frac{i}{2}\bar{\psi}_{\mu }\left( \delta ^{\left( 2\right) }\lambda
^{x}\right) .
\end{align}
}

\section{\label{4th-Order}Fourth Order}

Finally, at the fourth order we find\footnote{%
Note that $\nabla _{w}\nabla _{t}\nabla _{u}T^{xyz}=12\nabla _{w}\widetilde{E%
}^{xyz}{}_{tu}$ \cite{Cerchiai:2010xv}; similarly, $\nabla _{t}\nabla
_{z}\nabla _{y}h_{x}^{I}$ can be related to $\widetilde{E}$\textit{--tensor}
(\textit{cfr.} footnote 7).} {\footnotesize
\begin{align}
\left( \delta ^{\left( 4\right) }e_{\mu }^{a}\right) =& \frac{1}{2}\bar{%
\epsilon}\gamma ^{a}\left( \delta ^{\left( 3\right) }\psi _{\mu }\right) \,,
\\
\left( \delta ^{\left( 4\right) }\psi^i_{\mu }\right) =& \left( \delta
^{\left( 3\right) }{\mathcal{D}}_{\mu }\right) \epsilon^i -\frac{1}{6}%
\epsilon _{j}\bar{\lambda}^{ix}\gamma _{\mu }\left( \delta ^{\left( 3\right)
}\lambda _{x}^{j}\right) +\frac{1}{12}\gamma _{\mu \nu }\epsilon _{j}\bar{%
\lambda}^{ix}\gamma ^{\nu }\left( \delta ^{\left( 3\right) }\lambda
_{x}^{j}\right) +  \notag \\
& -\frac{1}{48}\gamma _{\mu \nu \rho }\epsilon _{j}\bar{\lambda}^{ix}\gamma
^{\nu \rho }\left( \delta ^{\left( 3\right) }\lambda _{x}^{j}\right) +\frac{1%
}{12}\gamma ^{\nu } \epsilon_j \bar{\lambda}^{ix}\gamma _{\mu \nu }\left(
\delta ^{\left( 3\right) }\lambda _{x}^{j}\right) +  \notag \\
& -\frac{1}{6}\epsilon _{j}\bar{\lambda}^{ix}\gamma _{a}\lambda
_{x}^{j}\left( \delta ^{\left( 3\right) }e_{\mu }^{a}\right) +  \notag \\
& -\frac{1}{3}\epsilon _{j}\left( \delta ^{\left( 3\right) }\bar{\lambda}%
^{ix}\right) \gamma _{\mu }\lambda _{x}^{j}+\frac{1}{12}\gamma _{ab}\epsilon
_{j}\bar{\lambda}^{ix}\gamma ^{b}\lambda _{x}^{j}\left( \delta ^{\left(
3\right) }e_{\mu }^{a}\right) +  \notag \\
& +\frac{1}{12}\gamma _{\mu \nu }\epsilon _{j}\left( \delta ^{\left(
3\right) }\bar{\lambda}^{ix}\right) \gamma ^{\nu }\lambda _{x}^{j}+  \notag
\\
& -\frac{1}{48}\gamma _{abc}\epsilon _{j}\bar{\lambda}^{ix}\gamma
^{bc}\lambda _{x}^{j}\left( \delta ^{\left( 3\right) }e_{\mu }^{a}\right) +
\notag \\
& -\frac{1}{48}\gamma _{\mu \nu \rho }\epsilon _{j}\left( \delta ^{\left(
3\right) }\bar{\lambda}^{ix}\right) \gamma ^{\nu \rho }\lambda _{x}^{j}+%
\frac{1}{12}\gamma ^{\nu }\epsilon _{j}\left( \delta ^{\left( 3\right) }\bar{%
\lambda}^{ix}\right) \gamma _{\mu \nu }\lambda _{x}^{j}+  \notag \\
& +\frac{1}{12}\gamma ^{b}\epsilon _{j}\bar{\lambda}^{ix}\gamma _{ab}\lambda
_{x}^{j}\left( \delta ^{\left( 3\right) }e_{\mu }^{a}\right) +  \notag \\
& +\frac{i}{4\sqrt{6}}h_{I}\tilde{F}_{\nu \rho }^{I}\left[ \left( \delta
^{\left( 3\right) }e_{\mu }^{a}\right) e_{b}^{\nu }e_{c}^{\rho }+e_{\mu
}^{a}\left( \delta ^{\left( 3\right) }e_{b}^{\nu }\right) e_{c}^{\rho
}+e_{\mu }^{a}e_{b}^{\nu }\left( \delta ^{\left( 3\right) }e_{c}^{\rho
}\right) +\right.  \notag \\
& +3e_{\mu }^{a}\left( \delta ^{\left( 2\right) }e_{b}^{\nu }\right) \left(
\delta ^{\left( 1\right) }e_{c}^{\rho }\right) +3e_{\mu }^{a}\left( \delta
^{\left( 1\right) }e_{b}^{\nu }\right) \left( \delta ^{\left( 2\right)
}e_{c}^{\rho }\right) +3\left( \delta ^{\left( 2\right) }e_{\mu }^{a}\right)
e_{b}^{\nu }\left( \delta ^{\left( 1\right) }e_{c}^{\rho }\right) +  \notag
\\
& +3\left( \delta ^{\left( 2\right) }e_{\mu }^{a}\right) \left( \delta
^{\left( 1\right) }e_{b}^{\nu }\right) e_{c}^{\rho }+3\left( \delta ^{\left(
1\right) }e_{\mu }^{a}\right) \left( \delta ^{\left( 2\right) }e_{b}^{\nu
}\right) e_{c}^{\rho }+3\left( \delta ^{\left( 1\right) }e_{\mu }^{a}\right)
e_{b}^{\nu }\left( \delta ^{\left( 2\right) }e_{c}^{\rho }\right) +  \notag
\\
& \left. +6\left( \delta ^{\left( 1\right) }e_{\mu }^{a}\right) \left(
\delta ^{\left( 1\right) }e_{b}^{\nu }\right) \left( \delta ^{\left(
1\right) }e_{c}^{\rho }\right) \right] \left( \gamma _{a}^{\phantom{a}%
bc}-4\delta _{a}^{b}\gamma ^{c}\right) \epsilon ^{i}+  \notag \\
& +\frac{i}{4}\nabla _{t}h_{Ix}\tilde{F}_{\nu \rho }^{I}\left[ \left( \delta
^{\left( 1\right) }e_{\mu }^{a}\right) e_{b}^{\nu }e_{c}^{\rho }+e_{\mu
}^{a}\left( \delta ^{\left( 1\right) }e_{b}^{\nu }\right) e_{c}^{\rho
}+\right.  \notag \\
& +\left. e_{\mu }^{a}e_{b}^{\nu }\left( \delta ^{\left( 1\right)
}e_{c}^{\rho }\right) \right] \left( \delta ^{\left( 1\right) }\phi
^{x}\right) \left( \delta ^{\left( 1\right) }\phi ^{t}\right) \left( \gamma
_{a}^{\phantom{a}bc}-4\delta _{a}^{b}\gamma ^{c}\right) \epsilon ^{i}+
\notag \\
& +\frac{i}{2}h_{Ix}\left( \delta ^{\left( 1\right) }\tilde{F}_{\nu \rho
}^{I}\right) \left[ \left( \delta ^{\left( 1\right) }e_{\mu }^{a}\right)
e_{b}^{\nu }e_{c}^{\rho }+e_{\mu }^{a}\left( \delta ^{\left( 1\right)
}e_{b}^{\nu }\right) e_{c}^{\rho }+\right.  \notag \\
& \left. +e_{\mu }^{a}e_{b}^{\nu }\left( \delta ^{\left( 1\right)
}e_{c}^{\rho }\right) \right] \left( \delta ^{\left( 1\right) }\phi
^{x}\right) \left( \gamma _{a}^{\phantom{a}bc}-4\delta _{a}^{b}\gamma
^{c}\right) \epsilon ^{i}+  \notag \\
& +\frac{i}{4}h_{Ix}\tilde{F}_{\nu \rho }^{I}\left[ \left( \delta ^{\left(
2\right) }e_{\mu }^{a}\right) e_{b}^{\nu }e_{c}^{\rho }+e_{\mu }^{a}\left(
\delta ^{\left( 2\right) }e_{b}^{\nu }\right) e_{c}^{\rho }+e_{\mu
}^{a}e_{b}^{\nu }\left( \delta ^{\left( 2\right) }e_{c}^{\rho }\right)
+\right.  \notag \\
& +2e_{\mu }^{a}\left( \delta ^{\left( 1\right) }e_{b}^{\nu }\right) \left(
\delta ^{\left( 1\right) }e_{c}^{\rho }\right) +2\left( \delta ^{\left(
1\right) }e_{\mu }^{a}\right) e_{b}^{\nu }\left( \delta ^{\left( 1\right)
}e_{c}^{\rho }\right)  \notag \\
& \left. +2\left( \delta ^{\left( 1\right) }e_{\mu }^{a}\right) \left(
\delta ^{\left( 1\right) }e_{b}^{\nu }\right) e_{c}^{\rho }\right] \left(
\delta ^{\left( 1\right) }\phi ^{x}\right) \left( \gamma _{a}^{\phantom{a}%
bc}-4\delta _{a}^{b}\gamma ^{c}\right) \epsilon ^{i}+  \notag \\
& +\frac{i}{4}h_{Ix}\tilde{F}_{\nu \rho }^{I}\left[ \left( \delta ^{\left(
1\right) }e_{\mu }^{a}\right) e_{b}^{\nu }e_{c}^{\rho }+e_{\mu }^{a}\left(
\delta ^{\left( 1\right) }e_{b}^{\nu }\right) e_{c}^{\rho }+e_{\mu
}^{a}e_{b}^{\nu }\left( \delta ^{\left( 1\right) }e_{c}^{\rho }\right) %
\right] \left( \delta ^{\left( 2\right) }\phi ^{x}\right) \left( \gamma
_{a}^{\phantom{a}bc}-4\delta _{a}^{b}\gamma ^{c}\right) \epsilon ^{i}+
\notag \\
& +\frac{i}{4}\sqrt{\frac{3}{2}}h_{I}\left( \delta ^{\left( 2\right) }\tilde{%
F}_{\nu \rho }^{I}\right) \left[ \left( \delta ^{\left( 1\right) }e_{\mu
}^{a}\right) e_{b}^{\nu }e_{c}^{\rho }+e_{\mu }^{a}\left( \delta ^{\left(
1\right) }e_{b}^{\nu }\right) e_{c}^{\rho }+e_{\mu }^{a}e_{b}^{\nu }\left(
\delta ^{\left( 1\right) }e_{c}^{\rho }\right) \right] \left( \gamma _{a}^{%
\phantom{a}bc}-4\delta _{a}^{b}\gamma ^{c}\right) \epsilon ^{i}+  \notag \\
& +\frac{i}{4}\sqrt{\frac{3}{2}}h_{I}\left( \delta ^{\left( 1\right) }\tilde{%
F}_{\nu \rho }^{I}\right) \left[ \left( \delta ^{\left( 2\right) }e_{\mu
}^{a}\right) e_{b}^{\nu }e_{c}^{\rho }+e_{\mu }^{a}\left( \delta ^{\left(
2\right) }e_{b}^{\nu }\right) e_{c}^{\rho }+e_{\mu }^{a}e_{b}^{\nu }\left(
\delta ^{\left( 2\right) }e_{c}^{\rho }\right) +\right.  \notag \\
& \left. +2e_{\mu }^{a}\left( \delta ^{\left( 1\right) }e_{b}^{\nu }\right)
\left( \delta ^{\left( 1\right) }e_{c}^{\rho }\right) +2\left( \delta
^{\left( 1\right) }e_{\mu }^{a}\right) e_{b}^{\nu }\left( \delta ^{\left(
1\right) }e_{c}^{\rho }\right) +2\left( \delta ^{\left( 1\right) }e_{\mu
}^{a}\right) \left( \delta ^{\left( 1\right) }e_{b}^{\nu }\right)
e_{c}^{\rho }\right] \left( \gamma _{a}^{\phantom{a}bc}-4\delta
_{a}^{b}\gamma ^{c}\right) \epsilon ^{i}+  \notag \\
& +\frac{i}{12}h_{Iz}\left( \delta ^{\left( 3\right) }\phi ^{z}\right)
\tilde{F}_{\nu \rho }^{I}\left( \gamma _{\mu }^{\phantom{\m}\nu \rho
}-4\delta _{\mu }^{\nu }\gamma ^{\rho }\right) \epsilon ^{i}+  \notag \\
& +\frac{i}{4\sqrt{6}}h_{I}\left( \delta ^{\left( 3\right) }\tilde{F}_{\nu
\rho }^{I}\right) \left( \gamma _{\mu }^{\phantom{\m}\nu \rho }-4\delta
_{\mu }^{\nu }\gamma ^{\rho }\right) \epsilon ^{i}+  \notag \\
& -\frac{1}{2}\left( \delta ^{\left( 1\right) }e_{\mu }^{a}\right) \epsilon
_{j}\bar{\lambda}^{ix}\gamma _{a}\left( \delta ^{\left( 2\right) }\lambda
_{x}^{j}\right) -\frac{1}{2}\epsilon _{j}\left( \delta ^{\left( 1\right) }%
\bar{\lambda}^{ix}\right) \gamma _{\mu }\left( \delta ^{\left( 2\right)
}\lambda _{x}^{j}\right) +  \notag \\
& +\frac{1}{4}\left( \delta ^{\left( 1\right) }e_{\mu }^{a}\right) \gamma
_{ab}\epsilon _{j}\bar{\lambda}^{ix}\gamma ^{b}\left( \delta ^{\left(
2\right) }\lambda _{x}^{j}\right) +\frac{1}{4}\gamma _{\mu \nu }\epsilon
_{j}\left( \delta ^{\left( 1\right) }\bar{\lambda}^{ix}\right) \gamma ^{\nu
}\left( \delta ^{\left( 2\right) }\lambda _{x}^{j}\right) +  \notag \\
& -\frac{1}{16}\left( \delta ^{\left( 1\right) }e_{\mu }^{a}\right) \gamma
_{abc}\epsilon _{j}\bar{\lambda}^{ix}\gamma ^{bc}\left( \delta ^{\left(
2\right) }\lambda _{x}^{j}\right) -\frac{1}{16}\gamma _{\mu \nu \rho
}\epsilon _{j}\left( \delta ^{\left( 1\right) }\bar{\lambda}^{ix}\right)
\gamma ^{\nu \rho }\left( \delta ^{\left( 2\right) }\lambda _{x}^{j}\right) +
\notag \\
& +\frac{1}{4}\gamma ^{b}\epsilon _{j}\bar{\lambda}^{ix}\gamma _{ab}\left(
\delta ^{\left( 2\right) }\lambda _{x}^{j}\right) \left( \delta ^{\left(
1\right) }e_{\mu }^{a}\right) +\frac{1}{4}\gamma ^{\nu }\epsilon _{j}\left(
\delta ^{\left( 1\right) }\bar{\lambda}^{ix}\right) \gamma _{\mu \nu }\left(
\delta ^{\left( 2\right) }\lambda _{x}^{j}\right) +  \notag \\
& -\frac{1}{2}\epsilon _{j}\bar{\lambda}^{ix}\gamma _{a}\left( \delta
^{\left( 1\right) }\lambda _{x}^{j}\right) \left( \delta ^{\left( 2\right)
}e_{\mu }^{a}\right) -\epsilon _{j}\left( \delta ^{\left( 1\right) }\bar{%
\lambda}^{ix}\right) \gamma _{a}\left( \delta ^{\left( 1\right) }\lambda
_{x}^{j}\right) \left( \delta ^{\left( 1\right) }e_{\mu }^{a}\right) +
\notag \\
& -\frac{1}{2}\epsilon _{j}\left( \delta ^{\left( 2\right) }\bar{\lambda}%
^{ix}\right) \gamma _{\mu }\left( \delta ^{\left( 1\right) }\lambda
_{x}^{j}\right) +\frac{1}{4}\left( \delta ^{\left( 2\right) }e_{\mu
}^{a}\right) \gamma _{ab} \epsilon_j \bar{\lambda}^{ix}\gamma ^{b}\left(
\delta ^{\left( 1\right) }\lambda _{x}^{j}\right) +  \notag \\
& +\frac{1}{2}\left( \delta ^{\left( 1\right) }e_{\mu }^{a}\right) \gamma
_{ab}\epsilon_j\left( \delta ^{\left( 1\right) }\bar{\lambda}^{ix}\right)
\gamma ^{b}\left( \delta ^{\left( 1\right) }\lambda _{x}^{j}\right) +\frac{1%
}{4}\gamma _{\mu \nu } \epsilon_j \left( \delta ^{\left( 2\right) }\bar{%
\lambda}^{ix}\right) \gamma ^{\nu }\left( \delta ^{\left( 1\right) }\lambda
_{x}^{j}\right) +  \notag \\
& -\frac{1}{16}\left( \delta ^{\left( 2\right) }e_{\mu }^{a}\right) \gamma
_{abc}\epsilon _{j}\bar{\lambda}^{ix}\gamma ^{bc}\left( \delta ^{\left(
1\right) }\lambda _{x}^{j}\right) -\frac{1}{8}\left( \delta ^{\left(
1\right) }e_{\mu }^{a}\right) \gamma _{abc}\epsilon _{j}\left( \delta
^{\left( 1\right) }\bar{\lambda}^{ix}\right) \gamma ^{bc}\left( \delta
^{\left( 1\right) }\lambda _{x}^{j}\right) +  \notag \\
& -\frac{1}{16}\gamma _{\mu \nu \rho }\epsilon _{j}\left( \delta ^{\left(
2\right) }\bar{\lambda}^{ix}\right) \gamma ^{\nu \rho }\left( \delta
^{\left( 1\right) }\lambda _{x}^{j}\right) +\frac{1}{4}\gamma ^{b}\epsilon
_{j}\bar{\lambda}^{ix}\gamma _{ab}\left( \delta ^{\left( 1\right) }\lambda
_{x}^{j}\right) \left( \delta ^{\left( 2\right) }e_{\mu }^{a}\right) +
\notag \\
& +\frac{1}{2}\gamma ^{b}\epsilon _{j}\left( \delta ^{\left( 1\right) }\bar{%
\lambda}^{ix}\right) \gamma _{ab}\left( \delta ^{\left( 1\right) }\lambda
_{x}^{j}\right) \left( \delta ^{\left( 1\right) }e_{\mu }^{a}\right) +\frac{1%
}{4}\gamma ^{\nu }\epsilon _{j}\left( \delta ^{\left( 2\right) }\bar{\lambda}%
^{ix}\right) \gamma _{\mu \nu }\left( \delta ^{\left( 1\right) }\lambda
_{x}^{j}\right) +  \notag \\
& -\frac{1}{2}\left( \delta ^{\left( 2\right) }e_{\mu }^{a}\right) \epsilon
_{j}\left( \delta ^{\left( 1\right) }\bar{\lambda}^{ix}\right) \gamma
_{a}\lambda _{x}^{j}-\frac{1}{2}\left( \delta ^{\left( 1\right) }e_{\mu
}^{a}\right) \epsilon _{j}\left( \delta ^{\left( 2\right) }\bar{\lambda}%
^{ix}\right) \gamma _{a}\lambda _{x}^{j}+  \notag \\
& +\frac{1}{4}\left( \delta ^{\left( 2\right) }e_{\mu }^{a}\right) \gamma
_{ab} \epsilon_j \left( \delta ^{\left( 1\right) }\bar{\lambda}^{ix}\right)
\gamma ^{b}\lambda _{x}^{j}+\frac{1}{4}\left( \delta ^{\left( 1\right)
}e_{\mu }^{a}\right) \epsilon_j \gamma _{ab}\left( \delta ^{\left( 2\right) }%
\bar{\lambda}^{ix}\right) \gamma ^{b}\lambda _{x}^{j}+  \notag \\
& -\frac{1}{16}\left( \delta ^{\left( 2\right) }e_{\mu }^{a}\right) \gamma
_{abc}\epsilon _{j}\left( \delta ^{\left( 1\right) }\bar{\lambda}%
^{ix}\right) \gamma ^{bc}\lambda _{x}^{j}-\frac{1}{16}\left( \delta ^{\left(
1\right) }e_{\mu }^{a}\right) \gamma _{abc}\epsilon _{j}\left( \delta
^{\left( 2\right) }\bar{\lambda}^{ix}\right) \gamma ^{bc}\lambda _{x}^{j}+
\notag \\
& +\frac{1}{4}\left( \delta ^{\left( 2\right) }e_{\mu }^{a}\right) \gamma
^{b} \epsilon_j \left( \delta ^{\left( 1\right) }\bar{\lambda}^{ix}\right)
\gamma _{ab}\lambda _{x}^{j} + \frac{1}{4}\left( \delta ^{\left( 1\right)
}e_{\mu }^{a}\right) \gamma ^{b} \epsilon_j \left( \delta ^{\left( 2\right) }%
\bar{\lambda}^{ix}\right) \gamma _{ab}\lambda _{x}^{j}+  \notag \\
& +\frac{i}{4}\left[ h_{Iz}\left( \delta ^{\left( 2\right) }\phi ^{z}\right)
+\nabla _{y}h_{Iz}\left( \delta ^{\left( 1\right) }\phi ^{z}\right) \left(
\delta ^{\left( 1\right) }\phi ^{y}\right) \right] \tilde{F}_{\nu \rho
}^{I}\left( \delta ^{\left( 1\right) }e_{\mu }^{a}\right) \left( \gamma
_{a}^{\phantom{a}\nu \rho }-4\delta _{a}^{\nu }\gamma ^{\rho }\right)
\epsilon ^{i}+  \notag \\
& +\frac{i}{12}\left[ \nabla _{y}h_{Iz}\left( \delta ^{\left( 1\right) }\phi
^{z}\right) \left( \delta ^{\left( 2\right) }\phi ^{y}\right) +2\nabla
_{y}h_{Iz}\left( \delta ^{\left( 2\right) }\phi ^{z}\right) \left( \delta
^{\left( 1\right) }\phi ^{y}\right) +\right.  \notag \\
& \left. +\nabla _{t}\nabla _{y}h_{Iz}\left( \delta ^{\left( 1\right) }\phi
^{z}\right) \left( \delta ^{\left( 1\right) }\phi ^{y}\right) \left( \delta
^{\left( 1\right) }\phi ^{t}\right) \right] \tilde{F}_{\nu \rho }^{I}\left(
\gamma _{\mu }^{\phantom{\m}\nu \rho }-4\delta _{\mu }^{\nu }\gamma ^{\rho
}\right) +  \notag \\
& +\frac{i}{4}\left[ h_{Iz}\left( \delta ^{\left( 2\right) }\phi ^{z}\right)
+\nabla _{y}h_{Iz}\left( \delta ^{\left( 1\right) }\phi ^{z}\right) \left(
\delta ^{\left( 1\right) }\phi ^{y}\right) \right] \left( \delta ^{\left(
1\right) }\tilde{F}_{\nu \rho }^{I}\right) \left( \gamma _{\mu }^{%
\phantom{\m}\nu \rho }-4\delta _{\mu }^{\nu }\gamma ^{\rho }\right) \epsilon
^{i}+  \notag \\
& +\frac{i}{4}h_{Iz}\left( \delta ^{\left( 1\right) }\phi ^{z}\right) \left(
\delta ^{\left( 2\right) }\tilde{F}_{\nu \rho }^{I}\right) \left( \gamma
_{\mu }^{\phantom{\m}\nu \rho }-4\delta _{\mu }^{\nu }\gamma ^{\rho }\right)
\epsilon ^{i} \,, \\
\left( \delta ^{\left( 4\right) }\phi ^{x}\right) =& \frac{i}{2}\bar{\epsilon%
}\left( \delta ^{\left( 3\right) }\lambda ^{x}\right) \,, \\
\left( \delta ^{\left( 4\right) }A_{\mu }^{I}\right) =& -\frac{1}{2}\bar{%
\epsilon}\gamma _{\mu }\left( \delta ^{\left( 3\right) }\lambda ^{x}\right)
h_{x}^{I}-\frac{1}{2}\left( \delta ^{\left( 3\right) }e_{\mu }^{a}\right)
\bar{\epsilon}\gamma _{a}\lambda ^{x}h_{x}^{I}+  \notag \\
& -\frac{i}{2}\sqrt{\frac{3}{2}}\bar{\epsilon}h^{I}\left( \delta ^{\left(
3\right) }\psi _{\mu }\right) +\frac{i}{2}h_{x}^{I}\left( \delta ^{\left(
3\right) }\phi ^{x}\right) \bar{\epsilon}\psi _{\mu }+  \notag \\
& -\frac{1}{2}\bar{\epsilon}\gamma _{\mu }\lambda ^{x}\nabla
_{y}h_{x}^{I}\left( \delta ^{\left( 3\right) }\phi ^{y}\right) +  \notag \\
& +\frac{3i}{2}h_{x}^{I}\left( \delta ^{\left( 1\right) }\phi ^{x}\right)
\bar{\epsilon}\left( \delta ^{\left( 2\right) }\psi _{\mu }\right) +\frac{3i%
}{2}h_{x}^{I}\left( \delta ^{\left( 2\right) }\phi ^{x}\right) \bar{\epsilon}%
\left( \delta ^{\left( 1\right) }\psi _{\mu }\right) +  \notag \\
& +\frac{3i}{2}\nabla _{y}h_{x}^{I}\left( \delta ^{\left( 1\right) }\phi
^{x}\right) \left( \delta ^{\left( 1\right) }\phi ^{y}\right) \bar{\epsilon}%
\left( \delta ^{\left( 1\right) }\psi _{\mu }\right) +\frac{i}{2}\nabla
_{y}h_{x}^{I}\left( \delta ^{\left( 2\right) }\phi ^{y}\right) \left( \delta
^{\left( 1\right) }\phi ^{x}\right) \bar{\epsilon}\psi _{\mu }+  \notag \\
& +i\nabla _{y}h_{x}^{I}\left( \delta ^{\left( 2\right) }\phi ^{x}\right)
\left( \delta ^{\left( 1\right) }\phi ^{y}\right) \bar{\epsilon}\psi _{\mu }+%
\frac{i}{2}\nabla _{z}\nabla _{y}h_{x}^{I}\left( \delta ^{\left( 1\right)
}\phi ^{x}\right) \left( \delta ^{\left( 1\right) }\phi ^{y}\right) \left(
\delta ^{\left( 1\right) }\phi ^{z}\right) \bar{\epsilon}\psi _{\mu }+
\notag \\
& -\frac{3}{2}\left( \delta ^{\left( 1\right) }e_{\mu }^{a}\right) \bar{%
\epsilon}\gamma _{a}\left( \delta ^{\left( 2\right) }\lambda ^{x}\right)
h_{x}^{I}-\frac{3}{2}\left( \delta ^{\left( 2\right) }e_{\mu }^{a}\right)
\bar{\epsilon}\gamma _{a}\left( \delta ^{\left( 1\right) }\lambda
^{x}\right) h_{x}^{I}+  \notag \\
& -\frac{3}{2}\bar{\epsilon}\gamma _{\mu }\left( \delta ^{\left( 2\right)
}\lambda ^{x}\right) \nabla _{y}h_{x}^{I}\left( \delta ^{\left( 1\right)
}\phi ^{y}\right) -3\left( \delta ^{\left( 1\right) }e_{\mu }^{a}\right)
\bar{\epsilon}\gamma _{a}\left( \delta ^{\left( 1\right) }\lambda
^{x}\right) \nabla _{y}h_{x}^{I}\left( \delta ^{\left( 1\right) }\phi
^{y}\right) +  \notag \\
& -\frac{3}{2}\left( \delta ^{\left( 2\right) }e_{\mu }^{a}\right) \bar{%
\epsilon}\gamma _{a}\lambda ^{x}\nabla _{y}h_{x}^{I}\left( \delta ^{\left(
1\right) }\phi ^{y}\right) -\frac{3}{2}\bar{\epsilon}\gamma _{\mu }\left(
\delta ^{\left( 1\right) }\lambda ^{x}\right) \left[ \nabla
_{y}h_{x}^{I}\left( \delta ^{\left( 2\right) }\phi ^{y}\right) +\right.
\notag \\
& \left. +\nabla _{z}\nabla _{y}h_{x}^{I}\left( \delta ^{\left( 1\right)
}\phi ^{y}\right) \left( \delta ^{\left( 1\right) }\phi ^{z}\right) \right] +
\notag \\
& -\frac{3}{2}\left( \delta ^{\left( 1\right) }e_{\mu }^{a}\right) \bar{%
\epsilon}\gamma _{a}\lambda ^{x}\left[ \nabla _{y}h_{x}^{I}\left( \delta
^{\left( 2\right) }\phi ^{y}\right) +\nabla _{z}\nabla _{y}h_{x}^{I}\left(
\delta ^{\left( 1\right) }\phi ^{y}\right) \left( \delta ^{\left( 1\right)
}\phi ^{z}\right) \right] +  \notag \\
& -\frac{1}{2}\bar{\epsilon}\gamma _{\mu }\lambda ^{x}\left[ \nabla
_{z}\nabla _{y}h_{x}^{I}\left( \delta ^{\left( 1\right) }\phi ^{y}\right)
\left( \delta ^{\left( 2\right) }\phi ^{z}\right) +2\nabla _{z}\nabla
_{y}h_{x}^{I}\left( \delta ^{\left( 2\right) }\phi ^{y}\right) \left( \delta
^{\left( 1\right) }\phi ^{z}\right) +\right.  \notag \\
& \left. +\nabla _{t}\nabla _{z}\nabla _{y}h_{x}^{I}\left( \delta ^{\left(
1\right) }\phi ^{y}\right) \left( \delta ^{\left( 1\right) }\phi ^{z}\right)
\left( \delta ^{\left( 1\right) }\phi ^{t}\right) \right] \,, \\
\left( \delta ^{\left( 4\right) }\lambda ^{ix}\right) =& -\frac{i}{2}\gamma
^{\mu }\left( \delta ^{\left( 3\right) }\widehat{\mathcal{D}}_{\mu }\phi
^{x}\right) \epsilon ^{i}-\frac{3i}{2}\gamma ^{a}\left( \delta ^{\left(
2\right) }e_{a}^{\mu }\right) \left( \delta ^{\left( 1\right) }\widehat{%
\mathcal{D}}_{\mu }\phi ^{x}\right) \epsilon ^{i}-\frac{3i}{2}\gamma
^{a}\left( \delta ^{\left( 1\right) }e_{a}^{\mu }\right) \left( \delta
^{\left( 2\right) }\widehat{\mathcal{D}}_{\mu }\phi ^{x}\right) \epsilon
^{i}+  \notag \\
& -\frac{i}{2}\gamma ^{a}\left( \delta ^{\left( 3\right) }e_{a}^{\mu
}\right) \widehat{\mathcal{D}}_{\mu }\phi ^{x}-\frac{1}{4\sqrt{6}}%
T^{xyz}\gamma ^{\mu }\epsilon _{j}\bar{\lambda}_{y}^{i}\gamma _{\mu }\left(
\delta ^{\left( 3\right) }\lambda _{z}^{j}\right) +  \notag \\
& +\frac{1}{4}\sqrt{\frac{3}{2}}T^{xyz}\epsilon _{j}\bar{\lambda}%
_{y}^{i}\left( \delta ^{\left( 3\right) }\lambda _{z}^{j}\right) -\frac{1}{8%
\sqrt{6}}T^{xyz}\gamma ^{\mu \nu }\epsilon _{j}\bar{\lambda}_{y}^{i}\gamma
_{\mu \nu }\left( \delta ^{\left( 3\right) }\lambda _{z}^{j}\right) +  \notag
\\
& +\frac{3}{4}\sqrt{\frac{3}{2}}T^{xyz}\epsilon _{j}\left( \delta ^{\left(
1\right) }\bar{\lambda}_{y}^{i}\right) \left( \delta ^{\left( 2\right)
}\lambda _{z}^{j}\right) -\frac{1}{8}\sqrt{\frac{3}{2}}T^{xyz}\gamma ^{\mu
\nu }\epsilon _{j}\left( \delta ^{\left( 1\right) }\bar{\lambda}%
_{y}^{i}\right) \gamma _{\mu \nu }\left( \delta ^{\left( 2\right) }\lambda
_{z}^{j}\right) +  \notag \\
& -\frac{1}{4}\sqrt{\frac{3}{2}}T^{xyz}\gamma ^{\mu }\epsilon _{j}\left(
\delta ^{\left( 1\right) }\bar{\lambda}_{y}^{i}\right) \gamma _{\mu }\left(
\delta ^{\left( 2\right) }\lambda _{z}^{j}\right) +  \notag \\
& -3\left( \delta ^{\left( 2\right) }\phi ^{y}\right) \Gamma _{yz}^{x}\left(
\delta ^{\left( 2\right) }\lambda ^{zi}\right) -3\left( \delta ^{\left(
3\right) }\phi ^{y}\right) \Gamma _{yz}^{x}\left( \delta ^{\left( 1\right)
}\lambda ^{zi}\right) +  \notag \\
& +\frac{3}{4}\sqrt{\frac{3}{2}}T^{xyz}\epsilon _{j}\left( \delta ^{\left(
2\right) }\bar{\lambda}_{y}^{i}\right) \left( \delta ^{\left( 1\right)
}\lambda _{z}^{j}\right) -\frac{1}{8}\sqrt{\frac{3}{2}}T^{xyz}\gamma ^{\mu
\nu }\epsilon _{j}\left( \delta ^{\left( 2\right) }\bar{\lambda}%
_{y}^{i}\right) \gamma _{\mu \nu }\left( \delta ^{\left( 1\right) }\lambda
_{z}^{j}\right) +  \notag \\
& -\frac{1}{4}\sqrt{\frac{3}{2}}T^{xyz}\gamma ^{\mu }\epsilon _{j}\left(
\delta ^{\left( 2\right) }\bar{\lambda}_{y}^{i}\right) \gamma _{\mu }\left(
\delta ^{\left( 1\right) }\lambda _{z}^{j}\right) +  \notag \\
& -\left( \delta ^{\left( 1\right) }\phi ^{y}\right) \Gamma _{yz}^{x}\left(
\delta ^{\left( 3\right) }\lambda ^{zi}\right) -\left( \delta ^{\left(
4\right) }\phi ^{y}\right) \Gamma _{yz}^{x}\lambda ^{zi}+  \notag \\
& +\frac{1}{4}\sqrt{\frac{3}{2}}T^{xyz}\epsilon _{j}\left( \delta ^{\left(
3\right) }\bar{\lambda}_{y}^{i}\right) \lambda _{z}^{j}-\frac{1}{8\sqrt{6}}%
T^{xyz}\gamma ^{\mu \nu }\epsilon _{j}\left( \delta ^{\left( 3\right) }\bar{%
\lambda}_{y}^{i}\right) \gamma _{\mu \nu }\lambda _{z}^{j}+  \notag \\
& -\frac{1}{4\sqrt{6}}T^{xyz}\gamma ^{\mu }\epsilon _{j}\left( \delta
^{\left( 3\right) }\bar{\lambda}_{y}^{i}\right) \gamma _{\mu }\lambda
_{z}^{j}+  \notag \\
& +\frac{3}{4}\sqrt{\frac{3}{2}}\nabla _{t}T^{xyz}\left( \delta ^{\left(
1\right) }\phi ^{t}\right) \epsilon _{j}\left( \delta ^{\left( 2\right) }%
\bar{\lambda}_{y}^{i}\right) \lambda _{z}^{j}+  \notag \\
& -\frac{1}{8}\sqrt{\frac{3}{2}}\nabla _{t}T^{xyz}\left( \delta ^{\left(
1\right) }\phi ^{t}\right) \gamma ^{\mu \nu }\epsilon _{j}\bar{\lambda}%
_{y}^{i}\gamma _{\mu \nu }\left( \delta ^{\left( 2\right) }\lambda
_{z}^{j}\right) +  \notag \\
& -\frac{1}{4}\sqrt{\frac{3}{2}}\nabla _{t}T^{xyz}\left( \delta ^{\left(
1\right) }\phi ^{t}\right) \gamma ^{\mu }\epsilon _{j}\bar{\lambda}%
_{y}^{i}\gamma _{\mu }\left( \delta ^{\left( 2\right) }\lambda
_{z}^{j}\right) +  \notag \\
& +\frac{3}{2}\sqrt{\frac{3}{2}}\nabla _{t}T^{xyz}\left( \delta ^{\left(
1\right) }\phi ^{t}\right) \epsilon _{j}\left( \delta ^{\left( 1\right) }%
\bar{\lambda}_{y}^{i}\right) \left( \delta ^{\left( 1\right) }\lambda
_{z}^{j}\right) +  \notag \\
& -\frac{1}{4}\sqrt{\frac{3}{2}}\nabla _{t}T^{xyz}\left( \delta ^{\left(
1\right) }\phi ^{t}\right) \gamma ^{\mu \nu }\epsilon _{j}\left( \delta
^{\left( 1\right) }\bar{\lambda}_{y}^{i}\right) \gamma _{\mu \nu }\left(
\delta ^{\left( 1\right) }\lambda _{z}^{j}\right) +  \notag \\
& -\frac{1}{2}\sqrt{\frac{3}{2}}\nabla _{t}T^{xyz}\left( \delta ^{\left(
1\right) }\phi ^{t}\right) \gamma ^{\mu }\epsilon _{j}\left( \delta ^{\left(
1\right) }\bar{\lambda}_{y}^{i}\right) \gamma _{\mu }\left( \delta ^{\left(
1\right) }\lambda _{z}^{j}\right) +  \notag \\
& +\frac{3}{4}\sqrt{\frac{3}{2}}\nabla _{t}T^{xyz}\left( \delta ^{\left(
1\right) }\phi ^{t}\right) \epsilon _{j}\left( \delta ^{\left( 2\right) }%
\bar{\lambda}_{y}^{i}\right) \lambda _{z}^{j}+  \notag \\
& -\frac{1}{8}\sqrt{\frac{3}{2}}\nabla _{t}T^{xyz}\left( \delta ^{\left(
1\right) }\phi ^{t}\right) \gamma ^{\mu \nu }\epsilon _{j}\left( \delta
^{\left( 2\right) }\bar{\lambda}_{y}^{i}\right) \gamma _{\mu \nu }\lambda
_{z}^{j}+  \notag \\
& -\frac{1}{4}\sqrt{\frac{3}{2}}\nabla _{t}T^{xyz}\left( \delta ^{\left(
1\right) }\phi ^{t}\right) \gamma ^{\mu }\epsilon _{j}\left( \delta ^{\left(
2\right) }\bar{\lambda}_{y}^{i}\right) \gamma _{\mu }\lambda _{z}^{j}+
\notag \\
& -6\left( \delta ^{\left( 2\right) }\phi ^{y}\right) \nabla _{t}\Gamma
_{yz}^{x}\left( \delta ^{\left( 1\right) }\phi ^{t}\right) \left( \delta
^{\left( 1\right) }\lambda ^{zi}\right) -3\left( \delta ^{\left( 1\right)
}\phi ^{y}\right) \nabla _{t}\Gamma _{yz}^{x}\left( \delta ^{\left( 1\right)
}\phi ^{t}\right) \left( \delta ^{\left( 2\right) }\lambda ^{zi}\right) +
\notag \\
& -3\left( \delta ^{\left( 3\right) }\phi ^{y}\right) \nabla _{t}\Gamma
_{yz}^{x}\left( \delta ^{\left( 1\right) }\phi ^{t}\right) \lambda ^{zi}+
\notag \\
& +\frac{3}{4}\sqrt{\frac{3}{2}}\epsilon _{j}\bar{\lambda}_{y}^{i}\left(
\delta ^{\left( 1\right) }\lambda _{z}^{j}\right) \left[ \nabla
_{t}T^{xyz}\left( \delta ^{\left( 2\right) }\phi ^{t}\right) +\nabla
_{u}\nabla _{t}T^{xyz}\left( \delta ^{\left( 1\right) }\phi ^{t}\right)
\left( \delta ^{\left( 1\right) }\phi ^{u}\right) \right] +  \notag \\
& -\frac{1}{8}\sqrt{\frac{3}{2}}\gamma ^{\mu \nu }\epsilon _{j}\bar{\lambda}%
_{y}^{i}\gamma _{\mu \nu }\left( \delta ^{\left( 1\right) }\lambda
_{z}^{j}\right) \left[ \nabla _{t}T^{xyz}\left( \delta ^{\left( 2\right)
}\phi ^{t}\right) +\nabla _{u}\nabla _{t}T^{xyz}\left( \delta ^{\left(
1\right) }\phi ^{t}\right) \left( \delta ^{\left( 1\right) }\phi ^{u}\right) %
\right] +  \notag \\
& -\frac{1}{4}\sqrt{\frac{3}{2}}\gamma ^{\mu }\epsilon _{j}\bar{\lambda}%
_{y}^{i}\gamma _{\mu }\left( \delta ^{\left( 1\right) }\lambda
_{z}^{j}\right) \left[ \nabla _{t}T^{xyz}\left( \delta ^{\left( 2\right)
}\phi ^{t}\right) +\nabla _{u}\nabla _{t}T^{xyz}\left( \delta ^{\left(
1\right) }\phi ^{t}\right) \left( \delta ^{\left( 1\right) }\phi ^{u}\right) %
\right] +  \notag \\
& +\frac{3}{4}\sqrt{\frac{3}{2}}\epsilon _{j}\left( \delta ^{\left( 1\right)
}\bar{\lambda}_{y}^{i}\right) \lambda _{z}^{j}\left[ \nabla
_{t}T^{xyz}\left( \delta ^{\left( 2\right) }\phi ^{t}\right) +\nabla
_{u}\nabla _{t}T^{xyz}\left( \delta ^{\left( 1\right) }\phi ^{t}\right)
\left( \delta ^{\left( 1\right) }\phi ^{u}\right) \right] +  \notag \\
& -\frac{1}{8}\sqrt{\frac{3}{2}}\gamma ^{\mu \nu }\epsilon _{j}\left( \delta
^{\left( 1\right) }\bar{\lambda}_{y}^{i}\right) \gamma _{\mu \nu }\lambda
_{z}^{j}\left[ \nabla _{t}T^{xyz}\left( \delta ^{\left( 2\right) }\phi
^{t}\right) +\nabla _{u}\nabla _{t}T^{xyz}\left( \delta ^{\left( 1\right)
}\phi ^{t}\right) \left( \delta ^{\left( 1\right) }\phi ^{u}\right) \right] +
\notag \\
& -\frac{1}{4}\sqrt{\frac{3}{2}}\gamma ^{\mu }\epsilon _{j}\left( \delta
^{\left( 1\right) }\bar{\lambda}_{y}^{i}\right) \gamma _{\mu }\lambda
_{z}^{j}\left[ \nabla _{t}T^{xyz}\left( \delta ^{\left( 2\right) }\phi
^{t}\right) +\nabla _{u}\nabla _{t}T^{xyz}\left( \delta ^{\left( 1\right)
}\phi ^{t}\right) \left( \delta ^{\left( 1\right) }\phi ^{u}\right) \right] +
\notag \\
& -3\left( \delta ^{\left( 1\right) }\phi ^{y}\right) \left[ \nabla
_{t}\Gamma _{yz}^{x}\left( \delta ^{\left( 2\right) }\phi ^{t}\right)
+\nabla _{u}\nabla _{t}\Gamma _{yz}^{x}\left( \delta ^{\left( 1\right) }\phi
^{t}\right) \left( \delta ^{\left( 1\right) }\phi ^{u}\right) \right] \left(
\delta ^{\left( 1\right) }\lambda ^{zi}\right) +  \notag \\
& -3\left( \delta ^{\left( 2\right) }\phi ^{y}\right) \left[ \nabla
_{t}\Gamma _{yz}^{x}\left( \delta ^{\left( 2\right) }\phi ^{t}\right)
+\nabla _{u}\nabla _{t}\Gamma _{yz}^{x}\left( \delta ^{\left( 1\right) }\phi
^{t}\right) \left( \delta ^{\left( 1\right) }\phi ^{u}\right) \right]
\lambda ^{zi}+  \notag \\
& +\frac{1}{4}\sqrt{\frac{3}{2}}\left[ \nabla _{u}T^{xyz}\left( \delta
^{\left( 3\right) }\phi ^{u}\right) +\nabla _{t}\nabla _{u}T^{xyz}\left(
\delta ^{\left( 1\right) }\phi ^{u}\right) \left( \delta ^{\left( 2\right)
}\phi ^{t}\right) +\right.  \notag \\
& \left. +2\nabla _{t}\nabla _{u}T^{xyz}\left( \delta ^{\left( 2\right)
}\phi ^{u}\right) \left( \delta ^{\left( 1\right) }\phi ^{t}\right) +\nabla
_{w}\nabla _{t}\nabla _{u}T^{xyz}\left( \delta ^{\left( 1\right) }\phi
^{u}\right) \left( \delta ^{\left( 1\right) }\phi ^{t}\right) \left( \delta
^{\left( 1\right) }\phi ^{w}\right) \right] \epsilon _{j}\bar{\lambda}%
_{y}^{i}\lambda _{z}^{j}+  \notag \\
& -\frac{1}{8\sqrt{6}}\gamma ^{\mu \nu }\epsilon _{j}\bar{\lambda}%
_{y}^{i}\gamma _{\mu \nu }\lambda _{z}^{j}\left[ \nabla _{u}T^{xyz}\left(
\delta ^{\left( 3\right) }\phi ^{u}\right) +\nabla _{t}\nabla
_{u}T^{xyz}\left( \delta ^{\left( 1\right) }\phi ^{u}\right) \left( \delta
^{\left( 2\right) }\phi ^{t}\right) +\right.  \notag \\
& \left. +2\nabla _{t}\nabla _{u}T^{xyz}\left( \delta ^{\left( 2\right)
}\phi ^{u}\right) \left( \delta ^{\left( 1\right) }\phi ^{t}\right) +\nabla
_{w}\nabla _{t}\nabla _{u}T^{xyz}\left( \delta ^{\left( 1\right) }\phi
^{u}\right) \left( \delta ^{\left( 1\right) }\phi ^{t}\right) \left( \delta
^{\left( 1\right) }\phi ^{w}\right) \right] +  \notag \\
& -\frac{1}{4\sqrt{6}}\gamma ^{\mu }\epsilon _{j}\bar{\lambda}_{y}^{i}\gamma
_{\mu }\lambda _{z}^{j}\left[ \nabla _{u}T^{xyz}\left( \delta ^{\left(
3\right) }\phi ^{u}\right) +\nabla _{t}\nabla _{u}T^{xyz}\left( \delta
^{\left( 1\right) }\phi ^{u}\right) \left( \delta ^{\left( 2\right) }\phi
^{t}\right) +\right.  \notag \\
& \left. +2\nabla _{t}\nabla _{u}T^{xyz}\left( \delta ^{\left( 2\right)
}\phi ^{u}\right) \left( \delta ^{\left( 1\right) }\phi ^{t}\right) +\nabla
_{w}\nabla _{t}\nabla _{u}T^{xyz}\left( \delta ^{\left( 1\right) }\phi
^{u}\right) \left( \delta ^{\left( 1\right) }\phi ^{t}\right) \left( \delta
^{\left( 1\right) }\phi ^{w}\right) \right] +  \notag \\
& -\left( \delta ^{\left( 1\right) }\phi ^{y}\right) \left[ \nabla
_{u}\Gamma _{yz}^{x}\left( \delta ^{\left( 3\right) }\phi ^{u}\right)
+\nabla _{u}\nabla _{t}\Gamma _{yz}^{x}\left( \delta ^{\left( 1\right) }\phi
^{t}\right) \left( \delta ^{\left( 2\right) }\phi ^{u}\right) +\right.
\notag \\
& +\left. 2\nabla _{u}\nabla _{t}\Gamma _{yz}^{x}\left( \delta ^{\left(
2\right) }\phi ^{t}\right) \left( \delta ^{\left( 1\right) }\phi ^{u}\right)
+\nabla _{w}\nabla _{u}\nabla _{t}\Gamma _{yz}^{x}\left( \delta ^{\left(
1\right) }\phi ^{t}\right) \left( \delta ^{\left( 1\right) }\phi ^{u}\right)
\left( \delta ^{\left( 1\right) }\phi ^{w}\right) \right] \lambda ^{zi}+
\notag \\
& +\frac{1}{4}\gamma \cdot \tilde{F}^{I}\left[ \nabla _{t}h_{I}^{x}\left(
\delta ^{\left( 3\right) }\phi ^{t}\right) +\nabla _{u}\nabla
_{t}h_{I}^{x}\left( \delta ^{\left( 1\right) }\phi ^{t}\right) \left( \delta
^{\left( 2\right) }\phi ^{u}\right) +\right.  \notag \\
& \left. +2\nabla _{u}\nabla _{t}h_{I}^{x}\left( \delta ^{\left( 2\right)
}\phi ^{t}\right) \left( \delta ^{\left( 1\right) }\phi ^{u}\right) \epsilon
^{i}+\nabla _{w}\nabla _{u}\nabla _{t}h_{I}^{x}\left( \delta ^{\left(
1\right) }\phi ^{t}\right) \left( \delta ^{\left( 1\right) }\phi ^{u}\right)
\left( \delta ^{\left( 1\right) }\phi ^{w}\right) \right] \epsilon ^{i}+
\notag \\
& +\frac{3}{4}\gamma \cdot \left( \delta ^{\left( 1\right) }\tilde{F}%
^{I}\right) \left[ \nabla _{t}h_{I}^{x}\left( \delta ^{\left( 2\right) }\phi
^{t}\right) +\nabla _{u}\nabla _{t}h_{I}^{x}\left( \delta ^{\left( 1\right)
}\phi ^{t}\right) \left( \delta ^{\left( 1\right) }\phi ^{u}\right) \right] +
\notag \\
& +\frac{3}{4}\gamma \cdot \left( \delta ^{\left( 2\right) }\tilde{F}%
^{I}\right) \nabla _{t}h_{I}^{x}\left( \delta ^{\left( 1\right) }\phi
^{t}\right) +\frac{1}{4}\gamma \cdot \left( \delta ^{\left( 3\right) }\tilde{%
F}^{I}\right) h_{I}^{x}+  \notag \\
& +\frac{1}{2}\gamma ^{ab}\left[ \left( \delta ^{\left( 3\right) }e_{a}^{\mu
}\right) e_{b}^{\nu } + 3 \left( \delta ^{\left( 2\right) }e_{a}^{\mu
}\right) \left( \delta ^{\left( 1\right) }e_{b}^{\nu }\right) \right] \tilde{%
F}_{\mu \nu }^{I}h_{I}^{x}+  \notag \\
& +\frac{3}{2}\gamma ^{ab}\left[ \left( \delta ^{\left( 2\right) }e_{a}^{\mu
}\right) e_{b}^{\nu } + \left( \delta ^{\left( 1\right) }e_{a}^{\mu }\right)
\left( \delta ^{\left( 1\right) }e_{b}^{\nu }\right) \right] \left( \delta
^{\left( 1\right) }\tilde{F}_{\mu \nu }^{I}\right) h_{I}^{x}+  \notag \\
& +\frac{3}{2}\gamma ^{ab}\left[ \left( \delta ^{\left( 2\right) }e_{a}^{\mu
}\right) e_{b}^{\nu } + \left( \delta ^{\left( 1\right) }e_{a}^{\mu }\right)
\left( \delta ^{\left( 1\right) }e_{b}^{\nu }\right) \right] \tilde{F}_{\mu
\nu }^{I}\nabla _{t}h_{I}^{x}\left( \delta ^{\left( 1\right) }\phi
^{t}\right) +  \notag \\
& +\frac{3}{2}\gamma ^{ab} \left( \delta ^{\left( 1\right) }e_{a}^{\mu
}\right) e_{b}^{\nu} \left( \delta ^{\left( 2\right) }\tilde{F}_{\mu \nu
}^{I}\right) h_{I}^{x}+  \notag \\
& +3 \gamma ^{ab} \left( \delta ^{\left( 1\right) }e_{a}^{\mu }\right)
e_{b}^{\nu} \left( \delta ^{\left( 1\right) }\tilde{F}_{\mu \nu }^{I}\right)
\nabla _{t}h_{I}^{x}\left( \delta ^{\left( 1\right) }\phi ^{t}\right) +
\notag \\
& +\frac{3}{2}\gamma ^{ab} \left( \delta ^{\left( 1\right) }e_{a}^{\mu
}\right) e_{b}^{\nu} \tilde{F}_{\mu \nu }^{I}\nabla _{t}\nabla
_{u}h_{I}^{x}\left( \delta ^{\left( 1\right) }\phi ^{t}\right) \left( \delta
^{\left( 1\right) }\phi ^{u}\right) +  \notag \\
& +\frac{3}{2}\gamma ^{ab} \left( \delta ^{\left( 1\right) }e_{a}^{\mu
}\right) e_{b}^{\nu} \tilde{F}_{\mu \nu }^{I}\nabla _{t}h_{I}^{x}\left(
\delta ^{\left( 1\right) }\phi ^{t}\right) \,,
\end{align}%
} with

{\footnotesize
\begin{align}
\left( \delta ^{\left( 3\right) }\tilde{F}_{\mu \nu }^{I}\right) =& \left(
\delta ^{\left( 3\right) }\mathcal{F}_{\mu \nu }^{I}\right) +3\left( \delta
^{\left( 1\right) }\bar{\psi}_{\left[ \mu \right. }\right) \gamma _{\left.
\nu \right] }\left( \delta ^{\left( 2\right) }\lambda ^{x}\right)
h_{x}^{I}+3\left( \delta ^{\left( 2\right) }\bar{\psi}_{\left[ \mu \right.
}\right) \gamma _{\left. \nu \right] }\left( \delta ^{\left( 1\right)
}\lambda ^{x}\right) h_{x}^{I}+  \notag \\
& +\bar{\psi}_{\left[ \mu \right. }\gamma _{\left. \nu \right] }\left(
\delta ^{\left( 3\right) }\lambda ^{x}\right) h_{x}^{I}+\left( \delta
^{\left( 3\right) }\bar{\psi}_{\left[ \mu \right. }\right) \gamma _{\left.
\nu \right] }\lambda ^{x}h_{x}^{I}+  \notag \\
& +\frac{i}{2}\sqrt{\frac{3}{2}}\bar{\psi}_{\mu }\left( \delta ^{\left(
3\right) }\psi _{\nu }\right) h^{I}+\frac{3i}{2}\sqrt{\frac{3}{2}}\left(
\delta ^{\left( 1\right) }\bar{\psi}_{\mu }\right) \left( \delta ^{\left(
2\right) }\psi _{\nu }\right) h^{I}+  \notag \\
& +\frac{3i}{2}\sqrt{\frac{3}{2}}\left( \delta ^{\left( 2\right) }\bar{\psi}%
_{\mu }\right) \left( \delta ^{\left( 1\right) }\psi _{\nu }\right) h^{I}+%
\frac{i}{2}\sqrt{\frac{3}{2}}\left( \delta ^{\left( 3\right) }\bar{\psi}%
_{\mu }\right) \psi _{\nu }h^{I}+  \notag \\
& -\frac{3i}{2}\bar{\psi}_{\mu }\left( \delta ^{\left( 2\right) }\psi _{\nu
}\right) h_{x}^{I}\left( \delta ^{\left( 1\right) }\phi ^{x}\right) -\frac{3i%
}{2}\bar{\psi}_{\mu }\left( \delta ^{\left( 1\right) }\psi _{\nu }\right)
h_{x}^{I}\left( \delta ^{\left( 2\right) }\phi ^{x}\right) +  \notag \\
& -3i\left( \delta ^{\left( 1\right) }\bar{\psi}_{\mu }\right) \left( \delta
^{\left( 1\right) }\psi _{\nu }\right) h_{x}^{I}\left( \delta ^{\left(
1\right) }\phi ^{x}\right) -\frac{3i}{2}\bar{\psi}_{\mu }\left( \delta
^{\left( 1\right) }\psi _{\nu }\right) \nabla _{y}h_{x}^{I}\left( \delta
^{\left( 1\right) }\phi ^{x}\right) \left( \delta ^{\left( 1\right) }\phi
^{y}\right) +  \notag \\
& -\frac{i}{2}\bar{\psi}_{\mu }\psi _{\nu }h_{x}^{I}\left( \delta ^{\left(
3\right) }\phi ^{x}\right) -\frac{3i}{2}\left( \delta ^{\left( 1\right) }%
\bar{\psi}_{\mu }\right) \psi _{\nu }h_{x}^{I}\left( \delta ^{\left(
2\right) }\phi ^{x}\right) +  \notag \\
& -\frac{3i}{2}\left( \delta ^{\left( 2\right) }\bar{\psi}_{\mu }\right)
\psi _{\nu }h_{x}^{I}\left( \delta ^{\left( 1\right) }\phi ^{x}\right) -%
\frac{i}{2}\bar{\psi}_{\mu }\psi _{\nu }\nabla _{y}h_{x}^{I}\left( \delta
^{\left( 1\right) }\phi ^{x}\right) \left( \delta ^{\left( 2\right) }\phi
^{y}\right) +  \notag \\
& -i\bar{\psi}_{\mu }\psi _{\nu }\nabla _{y}h_{x}^{I}\left( \delta ^{\left(
2\right) }\phi ^{x}\right) \left( \delta ^{\left( 1\right) }\phi ^{y}\right)
-\frac{3i}{2}\left( \delta ^{\left( 1\right) }\bar{\psi}_{\mu }\right) \psi
_{\nu }\nabla _{y}h_{x}^{I}\left( \delta ^{\left( 1\right) }\phi ^{x}\right)
\left( \delta ^{\left( 1\right) }\phi ^{y}\right) +  \notag \\
& -\frac{i}{2}\bar{\psi}_{\mu }\psi _{\nu }\nabla _{z}\nabla
_{y}h_{x}^{I}\left( \delta ^{\left( 1\right) }\phi ^{x}\right) \left( \delta
^{\left( 1\right) }\phi ^{y}\right) \left( \delta ^{\left( 1\right) }\phi
^{z}\right) +  \notag \\
& +3\bar{\psi}_{\left[ \mu \right. }\gamma _{\left. \nu \right] }\left(
\delta ^{\left( 1\right) }\lambda ^{x}\right) \nabla _{y}h_{x}^{I}\left(
\delta ^{\left( 2\right) }\phi ^{y}\right) +3\bar{\psi}_{\left[ \mu \right.
}\gamma _{\left. \nu \right] }\left( \delta ^{\left( 2\right) }\lambda
^{x}\right) \nabla _{y}h_{x}^{I}\left( \delta ^{\left( 1\right) }\phi
^{y}\right) +  \notag \\
& +6\left( \delta ^{\left( 1\right) }\bar{\psi}_{\left[ \mu \right. }\right)
\gamma _{\left. \nu \right] }\left( \delta ^{\left( 1\right) }\lambda
^{x}\right) \nabla _{y}h_{x}^{I}\left( \delta ^{\left( 1\right) }\phi
^{y}\right) +\bar{\psi}_{\left[ \mu \right. }\gamma _{\left. \nu \right]
}\lambda ^{x}\nabla _{y}h_{x}^{I}\left( \delta ^{\left( 3\right) }\phi
^{y}\right) +  \notag \\
& +3\left( \delta ^{\left( 1\right) }\bar{\psi}_{\left[ \mu \right. }\right)
\gamma _{\left. \nu \right] }\lambda ^{x}\nabla _{y}h_{x}^{I}\left( \delta
^{\left( 2\right) }\phi ^{y}\right) +3\left( \delta ^{\left( 2\right) }\bar{%
\psi}_{\left[ \mu \right. }\right) \gamma _{\left. \nu \right] }\lambda
^{x}\nabla _{y}h_{x}^{I}\left( \delta ^{\left( 1\right) }\phi ^{y}\right) +
\notag \\
& +3\bar{\psi}_{\left[ \mu \right. }\gamma _{\left. \nu \right] }\left(
\delta ^{\left( 1\right) }\lambda ^{x}\right) \nabla _{z}\nabla
_{y}h_{x}^{I}\left( \delta ^{\left( 1\right) }\phi ^{y}\right) \left( \delta
^{\left( 1\right) }\phi ^{z}\right) +  \notag \\
& +\bar{\psi}_{\left[ \mu \right. }\gamma _{\left. \nu \right] }\lambda
^{x}\nabla _{z}\nabla _{y}h_{x}^{I}\left( \delta ^{\left( 1\right) }\phi
^{y}\right) \left( \delta ^{\left( 2\right) }\phi ^{z}\right) +  \notag \\
& +2\bar{\psi}_{\left[ \mu \right. }\gamma _{\left. \nu \right] }\lambda
^{x}\nabla _{z}\nabla _{y}h_{x}^{I}\left( \delta ^{\left( 2\right) }\phi
^{y}\right) \left( \delta ^{\left( 1\right) }\phi ^{z}\right) +  \notag \\
& +3\left( \delta ^{\left( 1\right) }\bar{\psi}_{\left[ \mu \right. }\right)
\gamma _{\left. \nu \right] }\lambda ^{x}\nabla _{z}\nabla
_{y}h_{x}^{I}\left( \delta ^{\left( 1\right) }\phi ^{y}\right) \left( \delta
^{\left( 1\right) }\phi ^{z}\right) +  \notag \\
& +\bar{\psi}_{\left[ \mu \right. }\gamma _{\left. \nu \right] }\lambda
^{x}\nabla _{w}\nabla _{z}\nabla _{y}h_{x}^{I}\left( \delta ^{\left(
1\right) }\phi ^{y}\right) \left( \delta ^{\left( 1\right) }\phi ^{z}\right)
\left( \delta ^{\left( 1\right) }\phi ^{w}\right) +  \notag \\
& +3\left( \delta ^{\left( 2\right) }\bar{\psi}_{\left[ \mu \right. }\right)
\left( \delta ^{\left( 1\right) }e_{\left. \nu \right] }^{a}\right) \gamma
_{a}\lambda ^{x}h_{x}^{I}+3\left( \delta ^{\left( 1\right) }\bar{\psi}_{%
\left[ \mu \right. }\right) \left( \delta ^{\left( 2\right) }e_{\left. \nu %
\right] }^{a}\right) \gamma _{a}\lambda ^{x}h_{x}^{I}+  \notag \\
& +6\left( \delta ^{\left( 1\right) }\bar{\psi}_{\left[ \mu \right. }\right)
\left( \delta ^{\left( 1\right) }e_{\left. \nu \right] }^{a}\right) \gamma
_{a}\left( \delta ^{\left( 1\right) }\lambda ^{x}\right) h_{x}^{I}+6\left(
\delta ^{\left( 1\right) }\bar{\psi}_{\left[ \mu \right. }\right) \left(
\delta ^{\left( 1\right) }e_{\left. \nu \right] }^{a}\right) \gamma
_{a}\lambda ^{x}\nabla _{t}h_{x}^{I}\left( \delta ^{\left( 1\right) }\phi
^{t}\right) +  \notag \\
& +\bar{\psi}_{\left[ \mu \right. }\left( \delta ^{\left( 3\right)
}e_{\left. \nu \right] }^{a}\right) \gamma _{a}\lambda ^{x}h_{x}^{I}+3\bar{%
\psi}_{\left[ \mu \right. }\left( \delta ^{\left( 2\right) }e_{\left. \nu %
\right] }^{a}\right) \gamma _{a}\left( \delta ^{\left( 1\right) }\lambda
^{x}\right) h_{x}^{I}+  \notag \\
& +3\bar{\psi}_{\left[ \mu \right. }\left( \delta ^{\left( 2\right)
}e_{\left. \nu \right] }^{a}\right) \gamma _{a}\lambda ^{x}\nabla
_{t}h_{x}^{I}\left( \delta ^{\left( 1\right) }\phi ^{t}\right) +  \notag \\
& +3\bar{\psi}_{\left[ \mu \right. }\left( \delta ^{\left( 1\right)
}e_{\left. \nu \right] }^{a}\right) \gamma _{a}\left( \delta ^{\left(
2\right) }\lambda ^{x}\right) h_{x}^{I}+6\bar{\psi}_{\left[ \mu \right.
}\left( \delta ^{\left( 1\right) }e_{\left. \nu \right] }^{a}\right) \gamma
_{a}\left( \delta ^{\left( 1\right) }\lambda ^{x}\right) \nabla
_{t}h_{x}^{I}\left( \delta ^{\left( 1\right) }\phi ^{t}\right) +  \notag \\
& +3\bar{\psi}_{\left[ \mu \right. }\left( \delta ^{\left( 1\right)
}e_{\left. \nu \right] }^{a}\right) \gamma _{a}\lambda ^{x}\nabla _{u}\nabla
_{t}h_{x}^{I}\left( \delta ^{\left( 1\right) }\phi ^{t}\right) \left( \delta
^{\left( 1\right) }\phi ^{u}\right) +  \notag \\
& +3\bar{\psi}_{\left[ \mu \right. }\left( \delta ^{\left( 1\right)
}e_{\left. \nu \right] }^{a}\right) \gamma _{a}\lambda ^{x}\nabla
_{t}h_{x}^{I}\left( \delta ^{\left( 2\right) }\phi ^{t}\right) \,, \\
\left( \delta ^{\left( 3\right) }{\mathcal{D}}_{\mu }\right) =& \frac{1}{4}%
\left( \delta ^{\left( 3\right) }\omega _{\mu }^{ab}\right) \gamma _{ab} \,,
\\
\left( \delta ^{\left( 3\right) }\omega _{\mu }^{ab}\right) =& \frac{1}{2}%
\left( \delta ^{\left( 3 \right) } e_{c\mu }\right) \left( \Omega ^{abc} -
\Omega ^{bca} - \Omega ^{cab}\right) + 2 \left( \delta ^{\left( 2\right) }
e_{c\mu } \right) \left[ \left( \delta ^{\left( 1\right) }\Omega
^{abc}\right) -\left( \delta ^{\left( 1\right) }\Omega ^{bca}\right) -\left(
\delta ^{\left( 1\right) }\Omega ^{cab}\right) \right] +  \notag \\
&+ 2 \left( \delta ^{\left( 1\right) }e_{c\mu }\right) \left[ \left( \delta
^{\left( 2\right) }\Omega ^{abc}\right) -\left( \delta ^{\left( 2\right)
}\Omega ^{bca}\right) -\left( \delta ^{\left( 2\right) }\Omega ^{cab}\right) %
\right] +  \notag \\
& +\frac{1}{2} e_{c\mu} \left[ \left( \delta ^{\left( 3\right) }\Omega
^{abc}\right) -\left( \delta ^{\left( 3\right) }\Omega ^{bca}\right) -\left(
\delta ^{\left( 3\right) }\Omega ^{cab}\right) \right] +\left( \delta
^{\left( 3\right) }K_{\phantom{a}\mu }^{a\phantom{\mu}b}\right) \,, \\
\left( \delta ^{\left( 3\right) }\Omega ^{abc}\right) =& \left[ \left(
\delta ^{\left( 3\right) }e^{\mu a}\right) e^{\nu b} + 3 \left( \delta
^{\left( 2\right) }e^{\mu a}\right) \left( \delta ^{\left( 1\right) }e^{\nu
b}\right) + \right.  \notag \\
& \left. + 3 \left( \delta ^{\left( 1\right) }e^{\mu a}\right) \left( \delta
^{\left( 2\right) }e^{\nu b}\right) + e^{\mu a}\left( \delta ^{\left( 3
\right) }e^{\nu b}\right) \right] \left( \partial _{\mu }e_{\nu
}^{c}-\partial _{\nu }e_{\mu }^{c}\right) +  \notag \\
& + 3 \left[ \left( \delta ^{\left( 1\right) }e^{\mu a}\right) e^{\nu b} +
e^{\mu a}\left( \delta ^{\left( 1\right) }e^{\nu b}\right) \right] \left[
\partial _{\mu }\left( \delta ^{\left( 2\right) }e_{\nu }^{c}\right)
-\partial _{\nu }\left( \delta ^{\left( 2\right) }e_{\mu }^{c}\right) \right]
+  \notag \\
& + 3 \left[ \left( \delta ^{\left( 2\right) }e^{\mu a}\right) e^{\nu
b}+2\left( \delta ^{\left( 1\right) }e^{\mu a}\right) \left( \delta ^{\left(
1\right) }e^{\nu b}\right) +e^{\mu a}\left( \delta ^{\left( 2\right) }e^{\nu
b}\right) \right] \left[ \partial _{\mu }\left(\delta^{\left(1\right)}e_{\nu
}^{c}\right)- \partial _{\nu }\left(\delta^{\left(1\right)}e_{\mu
}^{c}\right)\right] +  \notag \\
& + e^{\mu a}e^{\nu b}\left[ \partial _{\mu }\left( \delta ^{\left( 3\right)
}e_{\nu }^{c}\right) -\partial _{\nu }\left( \delta ^{\left( 3\right)
}e_{\mu }^{c}\right) \right] \,, \\
\left(\delta^{\left(3\right)}K^a{}_\mu{}^b\right) = &\frac{3}{2}
\left(\delta^{\left(1\right)}\bar{\psi}_\rho\right)
\left(\delta^{\left(1\right)}e^{\rho a}\right) \gamma_c
\left(\delta^{\left(1\right)}e^c_{\mu}\right) \psi^b + \frac{3}{4}
\left(\delta^{\left(2\right)}\bar{\psi}_\rho\right) e^{\rho a} \gamma_c
\left(\delta^{\left(1\right)}e^c_\mu\right) \psi^b +  \notag \\
&+ \frac{3}{4} \left(\delta^{\left(2\right)}\bar{\psi}_\rho\right)
\left(\delta^{\left(1\right)}e^{\rho a}\right) \gamma_\mu \psi^b + \frac{3}{4%
} \left(\delta^{\left(1\right)}\bar{\psi}_\rho\right) e^{\rho a} \gamma_c
\left(\delta^{\left(2\right)}e^c_\mu\right) \psi^b +  \notag \\
&+ \frac{3}{4} \bar{\psi}_\rho \left(\delta^{\left(1\right)}e^{\rho
a}\right) \gamma_c \left(\delta^{\left(2\right)}e^c_\mu\right) \psi^b +
\frac{3}{4} \left(\delta^{\left(1\right)}\bar{\psi}_\rho\right)
\left(\delta^{\left(2\right)}e^{\rho a}\right) \gamma_\mu \psi^b +  \notag \\
& + \frac{3}{4} \bar{\psi}_\rho \left(\delta^{\left(2\right)}e^{\rho
a}\right) \gamma_c \left(\delta^{\left(1\right)}e^c_\mu\right) \psi^b +
\frac{1}{4} \left(\delta^{\left(3\right)}\bar{\psi}_\rho\right) e^{\rho a}
\gamma_\mu \psi^b +  \notag \\
&+ \frac{1}{4} \bar{\psi}^a \gamma_c
\left(\delta^{\left(3\right)}e^c_\mu\right) \psi^b + \frac{1}{4} \bar{\psi}%
_\rho \left(\delta^{\left(3\right)}e^{\rho a}\right) \gamma_\mu \psi^b +
\frac{1}{2} \bar{\psi}^{\left[a\right.} \gamma^{\left.b\right]}
\left(\delta^{\left(3\right)}\psi_\mu\right) +  \notag \\
& + \frac{1}{4} \bar{\psi}^a \gamma_\mu
\left(\delta^{\left(3\right)}\psi_\sigma\right) e^{\sigma b} + \frac{3}{2}
\left(\delta^{\left(1\right)}\bar{\psi}_\rho\right) e^{\rho \left[a\right.}
\gamma^{\left.b\right]} \left(\delta^{\left(2\right)}\psi_\mu\right) +
\notag \\
& + \frac{3}{4} \left(\delta^{\left(1\right)}\bar{\psi}_\rho\right) e^{\rho
a} \gamma_\mu \left(\delta^{\left(2\right)}\psi_\sigma\right) e^{\sigma b} +
\frac{3}{4} \bar{\psi}^a \gamma_c
\left(\delta^{\left(1\right)}e^c_\mu\right)
\left(\delta^{\left(2\right)}\psi_\sigma\right) e^{\sigma b} +  \notag \\
&+ \frac{3}{2} \bar{\psi}_\rho \left(\delta^{\left(1\right)}e^{\rho \left[%
a\right.}\right) \gamma^{\left.b\right]}
\left(\delta^{\left(2\right)}\psi_\mu\right) + \frac{3}{4} \bar{\psi}_\rho
\left(\delta^{\left(1\right)}e^{\rho a}\right) \gamma_\mu
\left(\delta^{\left(2\right)}\psi_\sigma\right) e^{\sigma b} +  \notag \\
&+ \frac{3}{4} \bar{\psi}^a \gamma_\mu
\left(\delta^{\left(2\right)}\psi_\sigma\right)
\left(\delta^{\left(1\right)}e^{\sigma b}\right) + \frac{3}{2}
\left(\delta^{\left(1\right)}\bar{\psi}_\rho\right) e^{\rho a} \gamma_c
\left(\delta^{\left(1\right)}e^c_\mu\right)
\left(\delta^{\left(1\right)}\psi_\sigma\right) e^{\sigma b} +  \notag \\
& + 3 \left(\delta^{\left(1\right)}\bar{\psi}_\rho\right)
\left(\delta^{\left(1\right)}e^{\rho \left[a\right.}\right) \gamma^{\left.b%
\right]} \left(\delta^{\left(1\right)}\psi_\mu\right) + \frac{3}{2}
\left(\delta^{\left(1\right)}\bar{\psi}_\rho\right)
\left(\delta^{\left(1\right)}e^{\rho a}\right) \gamma_\mu
\left(\delta^{\left(1\right)}\psi_\sigma\right) e^{\sigma b} +  \notag \\
& + \frac{3}{2} \bar{\psi}_\rho \left(\delta^{\left(1\right)}e^{\rho
a}\right) \gamma_c \left(\delta^{\left(1\right)}e^c_\mu\right)
\left(\delta^{\left(1\right)}\psi_\sigma\right) e^{\sigma b}+ \frac{3}{2}
\left(\delta^{\left(1\right)}\bar{\psi}_\rho\right) e^{\rho a} \gamma_\mu
\left(\delta^{\left(1\right)}\psi_\sigma\right)
\left(\delta^{\left(1\right)}e^{\sigma b}\right) +  \notag \\
&+ \frac{3}{2} \bar{\psi}^a \gamma_c
\left(\delta^{\left(1\right)}e^c_\mu\right)
\left(\delta^{\left(1\right)}\psi_\sigma\right)
\left(\delta^{\left(1\right)}e^{\sigma b}\right) + \frac{3}{2} \bar{\psi}%
_\rho \left(\delta^{\left(1\right)}e^{\rho a}\right) \gamma_\mu
\left(\delta^{\left(1\right)}\psi_\sigma\right)
\left(\delta^{\left(1\right)}e^{\sigma b}\right) +  \notag \\
& + \frac{3}{2} \left(\delta^{\left(2\right)}\bar{\psi}_\rho\right) e^{\rho %
\left[a\right.} \gamma^{\left.b\right]}
\left(\delta^{\left(1\right)}\psi_\mu\right) + \frac{3}{4}
\left(\delta^{\left(2\right)}\bar{\psi}_\rho\right) e^{\rho a} \gamma_\mu
\left(\delta^{\left(1\right)}\psi_\sigma\right) e^{\sigma b} +  \notag \\
& + \frac{3}{4} \bar{\psi}^a \gamma_c
\left(\delta^{\left(2\right)}e^c_\mu\right)
\left(\delta^{\left(1\right)}\psi_\sigma\right) e^{\sigma b} + \frac{3}{2}
\bar{\psi}_\rho \left(\delta^{\left(2\right)}e^{\rho \left[a\right.}\right)
\gamma^{\left.b\right]} \left(\delta^{\left(1\right)}\psi_\mu\right) +
\notag \\
& + \frac{3}{4} \bar{\psi}_\rho \left(\delta^{\left(2\right)}e^{\rho
a}\right) \gamma_\mu \left(\delta^{\left(1\right)}\psi_\sigma\right)
e^{\sigma b} + \frac{3}{4} \bar{\psi}^a \gamma_\mu
\left(\delta^{\left(1\right)}\psi_\sigma\right)
\left(\delta^{\left(2\right)}e^{\sigma b}\right) +  \notag \\
& + \frac{3}{2} \left(\delta^{\left(1\right)}\bar{\psi}_\rho\right) e^{\rho
a} \gamma_c \left(\delta^{\left(1\right)}e^c_\mu\right) \psi_\sigma
\left(\delta^{\left(1\right)}e^{\sigma b}\right) + \frac{3}{2}
\left(\delta^{\left(1\right)}\bar{\psi}_\rho\right)
\left(\delta^{\left(1\right)}e^{\rho a}\right) \gamma_\mu \psi_\sigma
\left(\delta^{\left(1\right)}e^{\sigma b}\right) +  \notag \\
& + \frac{3}{2} \bar{\psi}_\rho \left(\delta^{\left(1\right)}e^{\rho
a}\right) \gamma_c \left(\delta^{\left(1\right)}e^c_\mu\right) \psi_\sigma
\left(\delta^{\left(1\right)}e^{\sigma b}\right) + \frac{3}{2}
\left(\delta^{\left(2\right)}\bar{\psi}_\rho\right)
\left(\delta^{\left(1\right)}e^{\rho \left[a\right.}\right) \gamma^{\left.b%
\right]} \psi_\mu +  \notag \\
& + \frac{3}{4} \left(\delta^{\left(2\right)}\bar{\psi}_\rho\right) e^{\rho
a} \gamma_\mu \psi_\sigma \left(\delta^{\left(1\right)}e^{\sigma b}\right) +
\frac{3}{4} \bar{\psi}^a \gamma_c
\left(\delta^{\left(2\right)}e^c_\mu\right) \psi_\sigma
\left(\delta^{\left(1\right)}e^{\sigma b}\right) +  \notag \\
& + \frac{3}{2} \left(\delta^{\left(1\right)}\bar{\psi}_\rho\right)
\left(\delta^{\left(2\right)}e^{\rho \left[a\right.}\right) \gamma^{\left.b%
\right]} \psi_\mu + \frac{3}{4} \bar{\psi}_\rho
\left(\delta^{\left(2\right)}e^{\rho a}\right) \gamma_\mu \psi_\sigma
\left(\delta^{\left(1\right)}e^{\sigma b}\right) +  \notag \\
& + \frac{3}{4} \left(\delta^{\left(1\right)}\bar{\psi}_\rho\right) e^{\rho
a} \gamma_\mu \psi_\sigma \left(\delta^{\left(2\right)}e^{\sigma b}\right) +
\frac{3}{4} \bar{\psi}^a \gamma_c
\left(\delta^{\left(1\right)}e^c_\mu\right) \psi_\sigma
\left(\delta^{\left(2\right)}e^{\sigma b}\right) +  \notag \\
& + \frac{3}{4} \bar{\psi}_\rho \left(\delta^{\left(1\right)}e^{\rho
a}\right) \gamma_\mu \psi_\sigma \left(\delta^{\left(2\right)}e^{\sigma
b}\right) + \frac{1}{2} \left(\delta^{\left(3\right)}\bar{\psi}_\rho\right)
e^{\rho \left[a\right.} \gamma^{\left.b\right]} \psi_\mu +  \notag \\
& + \frac{1}{2} \bar{\psi}_\rho \left(\delta^{\left(3\right)}e^{\rho \left[%
a\right.}\right) \gamma^{\left.b\right]} \psi_\mu + \frac{1}{4} \bar{\psi}^a
\gamma_\mu \psi_\sigma \left(\delta^{\left(3\right)}e^{\sigma b}\right) \,,
\\
\left( \delta ^{\left( 3\right) }\widehat{{\mathcal{D}}}_{\mu }\phi
^{x}\right) =& \partial _{\mu }\left( \delta ^{\left( 3\right) }\phi
^{x}\right) -\frac{i}{2}\left( \delta ^{\left( 3\right) }\bar{\psi}_{\mu
}\right) \lambda ^{x}-\frac{3i}{2}\left( \delta ^{\left( 2\right) }\bar{\psi}%
_{\mu }\right) \left( \delta ^{\left( 1\right) }\lambda ^{x}\right) +  \notag
\\
& -\frac{3i}{2}\left( \delta ^{\left( 1\right) }\bar{\psi}_{\mu }\right)
\left( \delta ^{\left( 2\right) }\lambda ^{x}\right) -\frac{i}{2}\bar{\psi}%
_{\mu }\left( \delta ^{\left( 3\right) }\lambda ^{x}\right) .
\end{align}
}

\end{document}